\documentclass[a4paper,12pt]{article}
\usepackage[utf8]{inputenc}
 \usepackage{graphicx}
\usepackage{cancel}
\usepackage{amsmath, amssymb, amsfonts}
\usepackage{bbold}
\usepackage{latexsym}
\usepackage{subfigure}
\usepackage{authblk}
\usepackage{ dsfont }
\usepackage{ bbold }
\usepackage[font=small, format=plain, labelfont=bf,up, textfont=up]{caption}
\usepackage[top=3cm, bottom=3cm, left=2.5cm, right=2.5cm]{geometry}
\usepackage{setspace}
    \usepackage{sectsty}
    \sectionfont{\LARGE}
    \subsectionfont{\Large}
    \subsubsectionfont{\large}
\makeatletter
\def\maketag@@@#1{\hbox{\m@th\normalfont\normalsize#1}}
\makeatother

\usepackage{cite}

\usepackage[usenames,dvipsnames]{xcolor}
\usepackage{psfrag}
\usepackage{color, xcolor}
\usepackage{shadow}
\usepackage{feynarts}
\usepackage{fixltx2e}
\usepackage{mathrsfs}
\usepackage{fixltx2e}
\usepackage{tikz}
\usetikzlibrary{arrows,shapes}
\usetikzlibrary{trees}
\usetikzlibrary{matrix,arrows} 				
\usetikzlibrary{positioning}				
\usetikzlibrary{calc,through}				
\usetikzlibrary{decorations.pathreplacing}  
\usetikzlibrary{patterns}
\usepackage{pgffor}							

\usetikzlibrary{decorations.pathmorphing}	
\usetikzlibrary{decorations.markings}
\tikzset{
    vector/.style={decorate, decoration={snake}, draw},
	provector/.style={decorate, decoration={snake,amplitude=2.5pt}, draw},
	antivector/.style={decorate, decoration={snake,amplitude=-2.5pt}, draw},
    fermion/.style={draw=black, postaction={decorate},
        decoration={markings,mark=at position .55 with {\arrow[draw=black]{>}}}},
    fermionbar/.style={draw=black, postaction={decorate},
        decoration={markings,mark=at position .55 with {\arrow[draw=black]{<}}}},
    fermionnoarrow/.style={draw=black},
    gluon/.style={decorate, draw=black,
        decoration={coil,amplitude=4pt, segment length=5pt}},
    scalar/.style={dashed,draw=black, postaction={decorate},
        decoration={markings,mark=at position .55 with {\arrow[draw=black]{>}}}},
    scalarbar/.style={dashed,draw=black, postaction={decorate},
        decoration={markings,mark=at position .55 with {\arrow[draw=black]{<}}}},
    scalarnoarrow/.style={dashed,draw=black},
    electron/.style={draw=black, postaction={decorate},
        decoration={markings,mark=at position .55 with {\arrow[draw=black]{>}}}},
	bigvector/.style={decorate, decoration={snake,amplitude=4pt}, draw},
}

\tikzstyle{block} = [draw, rectangle, 
    minimum height=3em, minimum width=6em]
    
\usepackage[colorlinks=true,citecolor=Blue,urlcolor=Blue,linktocpage=true,
linkcolor=Blue]{hyperref}
\hypersetup{pageanchor=false}
\newcommand{\HBlabel}{Bechtle:2008jh,Bechtle:2011sb,Bechtle:2013gu,Bechtle:2013wla}
\newcommand{\FHlabel}{Heinemeyer:1998np,Heinemeyer:1998yj,Degrassi:2002fi,Heinemeyer:2007aq,Hahn:2013ria}

\newcommand{\bb}{b\bar{b}}
\newcommand{\tpm}{\tau^+\tau^-}
\newcommand{\tanb}{\tan\beta}
\newcommand{\br}{\textrm{BR}}
\newcommand{\fb}{\,\rm{fb} }

\newcommand{\ca}{c_{\alpha}}
\newcommand{\sa}{s_{\alpha}}
\newcommand{\cb}{c_{\beta}}
\newcommand{\sinb}{s_{\beta}}
\newcommand{\sw}{s_{W}}

\newcommand{\tb}{\tan\beta}

\newcommand{\sib}{s_{\beta}}
\newcommand{\sibn}{s_{\beta_n}}
\newcommand{\sibc}{s_{\beta_c}}
\newcommand{\cbn}{c_{\beta_n}}
\newcommand{\cbc}{c_{\beta_c}}

\newcommand{\ci}{\widetilde{\chi}_i^{0}}

\newcommand{\bpm}{\begin{pmatrix}}
\newcommand{\epm}{\end{pmatrix}}
\newcommand{\mhp}{M_{H^{\pm}}}

\newcommand{\psq}{p^{2}}
\newcommand{\ps}{(p^{2})}
\newcommand{\mmat}{\textbf{M}}
\newcommand{\gmat}{\hat{\boldsymbol{\Gamma}}}
\newcommand{\dmat}{\boldsymbol{\Delta}}
\newcommand{\gh}{\hat{\Gamma}_}

\newcommand{\seff}{\hat{\Sigma}^{\rm{eff}}_{ii}}
\newcommand{\dseff}{\hat{\Sigma}^{\rm{eff'}}_{ii}}

\newcommand{\Si}{\hat{\Sigma}_{ii}}

\newcommand{\Sij}{\hat{\Sigma}_{ij}}

\newcommand{\sig}{\hat{\Sigma}}
\newcommand{\Mm}{\mathcal{M}^{2}}
\newcommand{\Zb}{\hat{\textbf{Z}}}
\newcommand{\Zz}{\hat{Z}}
\newcommand{\Ub}{\textbf{U}}
\newcommand{\BW}{\Delta^{\textrm{BW}}}
\newcommand{\BWc}{\Delta^{\textrm{BW*}}}
\newcommand{\CP}{\mathcal{CP}}
\newcommand{\cmhmod}{\mathbb{C}M_h^{\rm{mod+}}}
\newcommand{\mhmod}{M_h^{\rm{mod+}}}

\newcommand{\gev}{\,\textrm{GeV}}
\newcommand{\tev}{\,\rm{TeV}}
\newcommand{\pat}{\phi_{A_t}}

\newcommand{\cp}{\mathcal{CP}}
\newcommand{\mphi}{M_{\phi}}
\newcommand{\lsim}{\lesssim}
\newcommand{\gsim}{\gtrsim}

\title{\hfill\small{\texttt{DESY 17-022}}\\ \vspace*{0.5cm}
\LARGE{
\textbf{Impact of $\cp$-violating interference effects on MSSM Higgs searches}
}}
\author{
\textsc{Elina Fuchs}$^{a,}$\textsuperscript{1},~~
\textsc{Georg Weiglein}$^{b,}$\textsuperscript{2}
\vspace{0.6cm}\\
\textit{(a) Department of Particle Physics and Astrophysics, }\\
 \textit{Weizmann Institute of Science, Rehovot 76100, Israel}\\
\textit{(b)  DESY, 
Notkestr. 85,
D-22607 Hamburg, Germany}
}
\footnotetext[1]{elina.fuchs@weizmann.ac.il}
\footnotetext[2]{georg.weiglein@desy.de}
\date{May 16, 2017}
\begin{document}

\maketitle
\thispagestyle{empty}
\begin{abstract}
Interference and mixing effects between neutral Higgs bosons in the MSSM
with complex parameters are shown to have a significant impact on the
interpretation of LHC 
searches for additional Higgs bosons.
Complex MSSM parameters introduce mixing between the $\cp$-even and
$\cp$-odd Higgs states 
and generate $\cp$-violating interference terms. Both
effects are enhanced in the case of almost degenerate states.
Employing as an example an extension of
a frequently used 
benchmark scenario by a non-zero phase $\pat$, the interference
contributions are obtained for the production 
of neutral Higgs bosons in gluon-fusion and in association with $b$-quarks
followed by the decay into a pair of $\tau$-leptons.
While the resonant mixing increases the individual cross sections
for the two heavy Higgs bosons $h_2$ and $h_3$,
strongly destructive interference effects between the contributions involving
$h_2$ and $h_3$
leave a considerable parameter region unexcluded that would appear to be
ruled out if the interference effects were neglected.
\end{abstract}
\newpage
\tableofcontents

\section{Introduction}
\label{chap:intro}
\setcounter{page}{1}

Despite a rich program of searches for additional Higgs bosons at the LHC,
no new scalars beyond the state at about 
$125\gev$ have been found so far.
Additional Higgs bosons may be hiding from the searches at the LHC where possible reasons can be reduced couplings, large masses or cancellations such as destructive interference effects.
 
Extended Higgs sectors are predicted by various models beyond the Standard
Model (SM) such as the Two-Higgs-Doublet Model (2HDM), the Minimal
Supersymmetric Standard Model (MSSM) and singlet extensions such as the
Next-to-Minimal Supersymmetric Standard Model 
(NMSSM). In this paper we focus on interpretations within the
MSSM.
The most obvious possibility in this model is to identify the 
discovered state with a mass of 
$M_h^{\rm{exp}}=125.09\pm0.24\gev$\,\cite{Aad:2015zhl} with the lightest
MSSM Higgs boson, which implies the existence of two heavier neutral Higgs
boson in the spectrum. 
The possibility that the discovered state could be the next-to-lightest
Higgs boson~\cite{Heinemeyer:2011aa,Bechtle:2012jw} is highly constrained
within the MSSM, but not ruled out (see Ref.~\cite{Bechtle:2016kui} for a 
recent update).

Although the properties of the discovered scalar are so far compatible with 
the ones predicted for the SM Higgs boson within the present experimental
uncertainties,
significant deviations from the SM are possible in individual Higgs
couplings, cross sections and branching ratios. Concerning the $\cp$
properties of the discovered state, a pure $\cp$-odd nature could be ruled
out~\cite{ATLAS:2015JP,Khachatryan:2014kca}, while only very weak bounds
exist so far on an admixture of $\cp$-even and $\cp$-odd components. 
If a non-zero $\cp$-odd admixture of the state at 125~GeV
could be experimentally established, this would be a direct 
manifestation of the presence of $\cp$ violation in the Higgs sector.
In general $\cp$ violation gives rise to a mixing
between all neutral Higgs bosons in the Higgs sector.
However, in the decoupling region of the MSSM the
lightest neutral Higgs boson is SM-like and almost purely $\cp$-even. 
On the other hand, the two heavy neutral Higgs bosons in this case
can very
significantly differ from $\cp$ eigenstates and have a large mixing
with each other. The search for additional heavy Higgs bosons should
therefore take into account the possibility that those heavy Higgs bosons
are not necessarily $\cp$ eigenstates. 

LHC searches for neutral MSSM Higgs bosons, following on from earlier
searches at LEP~\cite{Schael:2006cr} and the
Tevatron~\cite{Aaltonen:2009vf,Abazov:2010ci,Abazov:2011jh,Aaltonen:2011nh},
are carried out in the inclusive gluon fusion production channel,
$gg\rightarrow h_a~(a=1,2,3)$, and in the production process in association
with a pair of bottom quarks. 
For higher-order calculations of the cross sections of these two production processes within the MSSM see e.g.\ Refs.\,\cite{Dittmaier:2011ti,Dittmaier:2012vm,Heinemeyer:2013tqa,deFlorian:2016spz,Harlander:2012pb,Harlander:2016hcx,Harlander:2015xur} (assuming $\cp$ conservation in the Higgs sector) and \cite{Dedes:1999zh,Choi:1999aj,Choi:2001iu,Liebler:2016ceh} (including $\cp$ violation). 
The latter process, which dominates in the MSSM for high $\tan\beta$ values 
owing to an enhanced bottom Yukawa coupling, is usually written as 
$b\bar b\rightarrow h_a$ (according to the five-flavour scheme where the
bottom quark is regarded as a parton in the proton) or
$gg\rightarrow b\bar b h_a$ 
(according to the four-flavour scheme without a bottom parton density
distribution in the proton).
Searches for neutral MSSM Higgs bosons decaying to down-type fermions have
been carried out at the LHC in the
$\tau^{+}\tau^{-}$\,\cite{Chatrchyan:2011nx,Chatrchyan:2012vp,Khachatryan:2014wca,Aad:2014vgg,Aaboud:2016cre},
in the
$\mu^{+}\mu^{-}$\,\cite{CMS:2012lza,PERIEANU:2013vna,Thoma:2013mna,CMS:2015ooa}
and in the $b\bar b$\,\cite{Chatrchyan:2013qga,Khachatryan:2015tra} decay
channels,
where the $\tau^{+}\tau^{-}$ channel provides by far the highest sensitivity.
The decay modes 
of the heavy MSSM Higgs bosons
to down-type fermions are enhanced at large $\tan\beta$, whereas the
branching ratios of the heavy Higgs bosons into vector bosons vanish in the
decoupling limit\,\cite{Gunion:2002zf}. 

The results of the searches are on the one hand reported as nearly
model-independent limits on the product of the on-shell production cross
section (separately 
for the inclusive production in gluon fusion and the 
production in association with bottom quarks)
and the
considered branching ratio of a 
single scalar resonance,
assuming a narrow width. On the other hand, 
the search results are also interpreted in model-specific contexts using
appropriate benchmark scenarios,
for instance the $\mhmod$ scenario of the MSSM\,\cite{Carena:2013qia}.

In the MSSM the signal is potentially comprised of
contributions of all
three neutral Higgs bosons. The parameter region in which the masses 
of the two heavier neutral Higgs bosons of the MSSM are significantly
heavier
than the mass of the lightest neutral Higgs, where the latter needs to be
close to 125~GeV for a phenomenologically viable scenario, corresponds to
the decoupling region of the MSSM. The two heavier Higgs states are nearly
mass-degenerate over this whole region, with mass splittings that are often
below the experimental resolution. In the non-decoupling region even the
masses of all three neutral Higgs bosons can be close to each other.
In all model-specific interpretations at the LHC it has been assumed so far
that the signal contributions from different Higgs bosons can be added
incoherently, i.e.\ no interference contributions have been taken into
account. 
Under the assumption that $\cp$ is conserved in the Higgs sector, which is 
realised in the MSSM with real parameters, the mass eigenstates are states
of definite $\cp$, comprising the light and heavy $\cp$-even states $h$
and $H$, as well as the $\cp$-odd state $A$. In this case interference
effects between the heavy Higgs states $H$ and $A$ are absent, 
while the
$\cp$-even states $h$ and $H$ can interfere with each other.
If the
assumption of $\cp$ conservation in the Higgs sector is dropped,
interference effects occur also between the two heavy Higgs states, which
can be enhanced by a small mass difference between the two states. In the
MSSM $\cp$-violation in the Higgs sector is induced by potentially large 
loop corrections involving complex parameters. As a consequence,
in the general case of complex MSSM parameters
the $\cp$-even states $h$ and $H$ mix with the $\cp$-odd state $A$
into the mass eigenstates $h_1, h_2, h_3$.

The phenomenological consequences of
$\cp$-violating effects in the Higgs sector of
the MSSM have been investigated for the Higgs searches at 
LEP~\cite{Abbiendi:2004ww,Schael:2006cr}. 
In the analyses using 
the CPX benchmark scenario~\cite{Carena:2000ks} the
presence of a non-zero phase of the trilinear couplings $A_{t,b}$ and the
mixing between the $\cp$-even and $\cp$-odd states were found to have 
a significant impact on the limits, giving rise to unexcluded parameters
regions at much smaller values of the lightest Higgs mass than for the
$\cp$-conserving
case~\cite{Abbiendi:2004ww,Schael:2006cr,Williams:2007dc,Williams:2011bu}.

In the context of Higgs searches at the LHC,
interference effects 
between the light and the heavy neutral Higgs boson 
for the case of $\cp$-conservation in an extended Higgs sector
have been analysed in
Ref.\,\cite{Fuchs:2014ola} for Higgs production in the MSSM via the decay of
a heavy neutralino,
in Ref.\,\cite{Greiner:2015ixr} for $gg\rightarrow h/H
\rightarrow VV,\,V=W,Z$ within the 2HDM including background contributions,
and in Ref.\,\cite{Maina:2015ela} for the singlet extension of the SM. The
interference of two light NMSSM Higgs bosons decaying to two photons has been
investigated in Ref.\,\cite{Das:2017CPVNMSSM} for the $\cp$-violating case.
For
discussions of interference effects of heavy neutral Higgs bosons with each 
other and with
the background at low $\tan\beta$ in the $t\bar t$ final state see
Refs.\,\cite{Jung:2015gta,Bernreuther:2015fts,Carena:2016npr} 
and for interference with the background without $H-A$ interference see
Refs.~\cite{Quevillon:2016bod,ATLAS:2016pyq}

In the present paper we investigate the impact of mixing and 
interference effects on
the search for heavy Higgs bosons at the LHC in the channels 
$\left\lbrace \bb, gg \right\rbrace
\rightarrow h_1, h_2, h_3 \rightarrow \tau^+\tau^-$, where the production
modes denote the inclusive production in gluon fusion and the 
production in association with bottom quarks as discussed above.
In order to incorporate the interference effects we make use 
of a generalised 
narrow-width approximation (NWA)\,\cite{Fuchs:2014ola,Fuchs:2014zra}, where
in contrast to the standard NWA interference contributions are taken into
account. Our analysis is carried out in a benchmark scenario that is
extended such that it contains a non-zero phase of the 
trilinear coupling $A_t$. The exclusion bounds are evaluated with the help
of the program \texttt{HiggsBounds}\,\cite{\HBlabel}.
We find that the mixing effects between the two heavy Higgs bosons are 
resonantly enhanced in the parameter regions where the two Higgs bosons are
nearly mass-degenerate. The corresponding contributions of the 
intermediate Higgs states would yield a significant increase of the cross
section for the case of complex parameters as compared to the case of real
parameters if the interference contributions were neglected. However, the
interference effect between the two heavy Higgs bosons gives rise to a very
large destructive interference contribution, so that the net effect turns
out to be a large suppression of the total cross section times branching
ratio in the resonance region 
(see also Refs.~\cite{Fowler:2010eba,Fuchs:2015jwa,Fuchs:2016dyn}). 
As a consequence, a parameter region remains unexcluded by LHC searches from
Run~1 that would appear to be excluded if interference effects were neglected.

The paper is structured as follows. After fixing the notation for the MSSM
with complex parameters at tree level in Sect.\,\ref{sect:mssm}, we
summarise higher-order propagator-mixing in the Higgs sector and discuss the
mixing-enhancement of Higgs production cross sections in
Sect.\,\ref{sect:higherorder}. In Sect.\,\ref{sect:int}, we quantify
and investigate the interference contributions in the processes 
 $\left\lbrace \bb, gg \right\rbrace \rightarrow h_1, h_2, h_3 \rightarrow \tau^+\tau^-$.
Subsequently in Sect.\,\ref{sect:TheoExp}, we compare the predicted cross
sections with mixing and interference contributions to experimental limits
and evaluate their impact on the exclusion bounds in a benchmark
scenario
with a non-zero phase of the 
trilinear coupling $A_t$. 
Our conclusions are given in
Sect.\,\ref{sect:concl}.

\section{The MSSM with complex parameters at tree level}
\label{sect:mssm}
Before discussing higher-order mixing of Higgs bosons in Sect.\,\ref{sect:higherorder}, in this section we  will specify the notation for the MSSM with complex parameters at tree level, following Ref.\,\cite{Frank:2006yh}.

\paragraph{Sfermion sector}\label{sect:sfermion}
Sfermions $\tilde{f}_L, \tilde{f}_R$ mix into the mass eigenstates $\tilde{f}_1, \tilde{f}_2$ within one generation according to the mass matrix
\begin{align}
 M_{\tilde{f}}^{2}&=\left(
 \begin{matrix}
      M_{\tilde{f}_L}^{2}+m_f^{2}+M_Z^{2}\cos 2 \beta (I_f^{3}-Q_f\sw^{2}) & m_f X_f^{*}\\
      m_f X_f & M_{\tilde{f}_R}^{2}+m_{f}^{2}+M_Z^{2}\cos2\beta Q_f\sw^{2}
                        \end{matrix} \right)\,,\label{eq:Msf}
\end{align}
where $X_f := A_f - \mu^{*}\cdot \left\{\cot \beta,  \tan \beta\right\}$ for $f$ being an up- or down-type quark, respectively.
Besides the trilinear couplings $A_f=|A_f|e^{i\phi_{A_f}}$ also the higgsino mass parameter $\mu=|\mu|e^{i\phi_{\mu}}$ can be complex. Starting from one-loop order, these phases may influence the Higgs sector via sfermion loops.

\paragraph{Gluino sector}
The mass of a gluino $\tilde{g}^{a},~a=1,2,3,$ is given by 
$ m_{\tilde{g}}=|M_3|$,
where $M_3=|M_3|\,e^{i\phi_{M_3}}$ is the possibly complex gluino mass parameter. 
The gluino does not directly couple to Higgs bosons. Hence, the phase
$\phi_{M_3}$ enters the 
predictions for the Higgs-boson masses and the wave function normalisation
factors for external Higgs bosons only from the two-loop level onwards
whereas it has an impact for example on the bottom Yukawa coupling already
at one-loop order. 

\paragraph{Neutralino and chargino sector}
\label{sect:neucha}
The charginos $\widetilde{\chi}_i^{\pm},~i=1,2$, are admixtures of the charged winos $\widetilde{W}^{\pm}$ and higgsinos $\widetilde{H}^{\pm}$ via the mass matrix
\begin{equation}
  X =  \bpm M_2 & \sqrt{2}M_W\sib\\ \sqrt{2}M_W \cb & \mu \epm.
  \label{eq:X}
\end{equation}
Likewise the neutralinos $\ci,\,i=1,...,4$, are composed of the neutral electroweak gauginos $\widetilde{B},\,\widetilde{W}^{3}$ and the neutral Higgsinos $\tilde{h}_d^0,\, \tilde{h}_u^0$:
\begin{equation}
Y = \left(\begin{matrix}
	M_1 & 0 & -M_Z \cb s_W & M_Z \sinb s_W\\
	0 & M_2 & M_Z \cb c_W & -M_Z \sinb c_W \\
	-M_Z \cb s_W & M_Z \cb c_W & 0 &-\mu\\
	M_Z \sinb s_W & -M_Z \sinb c_W &  -\mu &0\\
	     \end{matrix} \right) .
\label{eq:neutmix}
\end{equation}
Thus, at tree-level, mixing in the chargino sector is governed by the higgsino and wino mass parameters $\mu$ and $M_2$, respectively, and in the neutralino sector in addition by the bino mass parameter $M_1$.
Although all of these three parameters can be complex in principle, only two of 
the phases are independent, and the choice $\phi_{M_2}=0$ is a common convention.

\paragraph{Higgs sector}\label{sect:Higgstree}
The two complex scalar Higgs doublets of the MSSM are denoted as
\begin{align}
 \mathcal{H}_1 &= \begin{pmatrix}h_d^{0}\\h_d^{-}\end{pmatrix}
                  =\bpm v_d+\frac{1}{\sqrt{2}}(\phi_1^{0}-i\chi_1^{0})\\ -\phi_1^{-}\epm \label{eq:H1doublet}\,,\\
 \mathcal{H}_2 &= \begin{pmatrix}h_u^{+}\\h_u^{0}\end{pmatrix}
                  = \bpm \phi_2^{+}\\v_u+\frac{1}{\sqrt{2}}(\phi_2^{0}+i\chi_2^{0})\epm \label{eq:H2doublet}\,.
\end{align}
As a possible relative phase between both doublets
vanishes at the minimum of the 
Higgs potential and the phase of the coefficient of the bilinear term 
in the Higgs potential can be rotated away, 
the Higgs sector is $\CP$ conserving at lowest order. The tree level mass eigenstates as eigenstates of $\cp$ result from the diagonalisation of the mass matrices of the neutral and the charged components, respectively,
\begin{align}
 \bpm h\\H\\A\\G \epm = \bpm -\sa& \ca & 0 & 0\\ \ca &\sa&0&0\\ 0&0& -\sibn &\cbn \\ 0&0 & \cbn& \sibn \epm \bpm \phi_1^{0}\\ \phi_2^{0}\\ \chi_1^{0}\\ \chi_2^{0}\epm,\hspace*{1cm}
 \bpm H^{\pm}\\G^{\pm}\epm = \bpm  -\sibc& \cbc\\ \cbc &\sibc\epm 
 \bpm \phi_1^{\pm}\\ \phi_2^{\pm}\epm\label{eq:ncHmix},
\end{align}
with the short-hand notation $s_x\equiv\sin x,~c_x\equiv\cos x$. 
The mixing angle $\alpha$ 
is associated with the $\mathcal{CP}$-even Higgs bosons $h,H$; $\beta_n$
with the 
neutral $\mathcal{CP}$-odd Higgs $A$ and Goldstone boson $G$, and $\beta_c$
with 
the charged Higgs $H^{\pm}$ and the charged Goldstone boson $G^{\pm}$. 
The masses of the $\mathcal{CP}$-odd and the charged Higgs bosons are at tree level related by
 $m_{H^{\pm}}^{2} = m_A^{2} + M_W^{2}$.
At lowest order, the Higgs sector is fully determined by the two SUSY 
input parameters (in addition to SM masses and gauge couplings) $\tan\beta\equiv v_u/v_d$ and 
$m_{H^{\pm}}$ (or, for conserved $\mathcal{CP}$, equivalently $m_A$).
We use in this paper a lower-case and upper-case notation in
order to indicate tree-level and loop-corrected masses, respectively.

In the \textit{decoupling limit} of $M_A\gg M_Z$ 
the MSSM Higgs sector appears SM-like,
and the heavy Higgs bosons are difficult to detect
in production and decay channels involving gauge bosons.
On the other hand, the couplings to fermions can be either suppressed or enhanced, depending on the angles $\alpha$ and $\beta$.
In the processes considered in this work, in particular the couplings of the neutral Higgs bosons $i=h,H,A$ to down-type fermions $f_d$ such as $\tau$-leptons and $b$-quarks are involved\,\cite{Gunion:1989we},
\begin{equation}
 g_{i,f_d\bar{f}_d}^{\rm{tree}} = -\frac{igm_{f_d}}{2M_W} \,\cdot \left\{-\frac{\sa}{\cb},~\frac{\ca}{\cb},~ i\gamma_5\tb \right\},~~i=h,H,A\,.\label{eq:GHtt}
\end{equation}
As $\ca\rightarrow \sinb$ in the decoupling limit, the couplings of $H$ and $A$ to down-type fermions are enhanced by large values of $\tb$.

\section{Higher-order $\cp$-violating Higgs mixing}
\label{sect:higherorder}
Higher-order corrections have a sizeable impact on the MSSM Higgs sector.
In loop diagrams, particles from all other sectors contribute to Higgs observables such that  
in particular the trilinear couplings $A_f$, the stop and sbottom masses, the gluino mass, and in 
the sub-leading terms the higgsino mass parameter $\mu$, play -- besides $M_{H^{\pm}}$ (or $M_A$) and $\tan\beta$ as at lowest order -- an important role in the Higgs phenomenology. 
While $\cp$ is conserved in the Higgs sector at lowest order, those
parameters from other sectors that are a priori complex can introduce $\cp$
violation in the Higgs sector at higher orders. In the $\cp$-conserving
case, there is only a $2\times 2$ mixing among the two $\cp$-even neutral
Higgs bosons, $h$ and $H$. On the contrary, the non-zero phases of complex
parameters cause a $\cp$-violating mixing of the scalars $h, H$ and the
pseudoscalar $A$ into the mass eigenstates $h_1, h_2, h_3$ (in addition,
there is also mixing with the neutral Goldstone and vector bosons that we
will neglect in this work, see Ref.~\cite{Frank:2006yh} for a discussion). 
This $3\times 3$ mixing structure is reflected in the mass matrix of the $h,
H, A$ system and in the on-shell wave-function normalisation factors, $\Zb$,
as we will outline in Sect.\,\ref{sect:propZ} following
Refs.\,\cite{Frank:2006yh,Williams:2011bu,Fuchs:2016swt}. In
Sect.\,\ref{sect:mixenhance} we will discuss
the consequences of highly admixed mass eigenstates on the production cross sections of each resonance.

For the calculation of radiative corrections, we adopt the hybrid on-shell
and $\overline{\rm{DR}}$-renormalisation scheme defined in
Ref.\,\cite{Frank:2006yh} such that the masses are renormalised on-shell
whereas $\overline{\rm{DR}}$ conditions are employed for the fields and
$\tan\beta$. We evaluate the Higgs masses, widths, branching ratios and
self-energies with \texttt{FeynHiggs-2.10.2}\,\cite{\FHlabel}\footnote{The
additional corrections included in more recent versions do not qualitatively
alter the effects discussed in this paper. The latest version will be
used in an update for the Run 2 results of the LHC
searches~\cite{IntCalc:InProgress}.}, including the full momentum-dependent
one-loop corrections and leading two-loop contributions.

\subsection{Propagator corrections and the $\Zb$-matrix}
\label{sect:propZ}
\paragraph{Propagator matrix}
In the general case, all renormalised self-energies $\Sij\ps$  of the Higgs bosons $i,j=h,H,A$ are non-zero. Hence, the mass-square matrix $\mmat$ of the neutral Higgs bosons  does not only consist of the tree-level masses-squares $m_i^2$ in the diagonal entries, but also of the momentum-dependent renormalised self-energies $\Sij\ps$ on the diagonal and off-diagonal positions. Therefore the matrix of mass-squares becomes
\begin{equation}
 \mmat(\psq) = \bpm
  m_h^{2}-\hat{\Sigma}_{hh}(\psq) & -\hat{\Sigma}_{hH}(\psq) &-\hat{\Sigma}_{hA}(\psq)\\
  -\hat{\Sigma}_{Hh}(\psq) & m_H^{2}-\hat{\Sigma}_{HH}(\psq) &-\hat{\Sigma}_{HA}(\psq)\\
  -\hat{\Sigma}_{Ah}(\psq) & -\hat{\Sigma}_{AH}(\psq) &m_A^{2}-\hat{\Sigma}_{AA}(\psq)
\epm \label{eq:HiggsMassMatrix}\,.
\end{equation}
The self-energies $\Sij\psq$ also contribute to the renormalised irreducible two-point vertex functions,
\begin{equation}
 \hat{\Gamma}_{ij}\ps = i\left[(\psq-m_i^{2})\delta_{ij}+\sig_{ij}\ps\right] \label{eq:IrepVert}\,,
\end{equation}
whose elements form the $3\times3$ matrix $\gmat_{hHA}\ps$ that is related to the mass-square matrix $\mmat\ps$ and to the propagator matrix $\dmat\ps$ via
\begin{equation}
 \dmat_{hHA}(\psq) = - \left[\gmat_{hHA}(\psq) \right]^{-1}
 = - \left[  i\left(\psq \textbf{1} -\mmat(\psq) \right) \right]^{-1}\,. \label{eq:DeltaGamInv}
\end{equation}
The elements of $\dmat\ps$ are the propagators $\Delta_{ij}\ps$ starting as Higgs boson $i$ and ending on $j$ with all possible mixings in between. The off-diagonal propagators ($i\neq j$) have the form
\begin{equation}
 \Delta_{ij}(\psq) = \frac{\gh{ij}\gh{kk}- \gh{jk}\gh{ki}}{\gh{ii}\gh{jj}\gh{kk} +2\gh{ij}\gh{jk}\gh{ki}-\gh{ii}\gh{jk}^{2} -\gh{jj}\gh{ki}^{2}-\gh{kk}\gh{ij}^{2}} \label{eq:Dij}\,,
\end{equation}
where the explicit $\psq$-dependence of all $\hat{\Gamma}\ps$ has been
suppressed in order to simplify the notation. The diagonal propagator can be expressed as
\begin{align}
 \Delta_{ii}\ps &= \frac{\gh{jj}\gh{kk}-\gh{jk}^{2}}{-\gh{ii}\gh{jj}\gh{kk} + \gh{ii}\gh{jk}^{2} - 2\gh{ij}\gh{jk}\gh{ki} + \gh{jj}\gh{ki}^{2} +\gh{kk}\gh{ij}^{2}}\label{eq:Dii}\\
&= \frac{i}{\psq -m_i^{2} + \seff\ps}\,,\label{eq:Diieff}
\end{align}
where in the last step the effective self-energy $\seff$ is defined as
\begin{align}
   \seff(p^{2}) &= \hat{\Sigma}_{ii}(p^{2}) - i \frac{2\gh{ij}(p^{2})\gh{jk}(p^{2})\gh{ki}(p^{2}) - \gh{ki}^{2}(\psq)\gh{jj}(\psq)-\gh{ij}^{2}(\psq)\gh{kk}(\psq)}{\gh{jj}(\psq)\gh{kk}(\psq)-\gh{jk}^{2}(\psq)} \label{eq:seff}\,,
\end{align}
starting with the diagonal self-energy $\Si$ at one-loop 
order and containing the mixing 2-point functions whose products in $\seff$ arise from two-loop order onwards. 
The complex poles $\Mm_{a}$, where $a = 1, 2, 3$ label the mass eigenstates, are determined as
solutions of the equations
\begin{align}
\psq -m_i^{2} + \seff\ps = 0 ,\label{eq:complexpole}
\end{align}
with $i = h, H, A$.

\paragraph{Z-factors}
For
the correct normalisation of the $S$-matrix elements appropriate on-shell
properties of all external particles are required. Pure on-shell
renormalisation schemes guarantee these properties, whereas the
$\overline{\rm{DR}}$ field renormalisation conditions make the finite
on-shell wave-function normalisation factors necessary which ensure that
external particles have unit residue and that all mixing contributions
vanish on-shell.
According to the LSZ formalism\,\cite{Lehmann:1954rq}, they are obtained
from the residue of the propagator of the external field, here a neutral
Higgs boson $i=h,H,A$, evaluated at one of the complex poles
$\Mm_a,~a=1,2,3$\,\cite{Dabelstein:1995js,Heinemeyer:2001iy,Williams:2007dc,Williams:2011bu,Fuchs:2016swt},
 \begin{align}
 \Zz_i^{a}
 & =\frac{1}{1+\dseff(\Mm_a)} \label{eq:Zha}\,.
\end{align}
The normalisation of each external particle $i$ at the mass shell $\psq=\Mm_a$ is accounted for by a factor of $\sqrt{\Zz_i^{a}}$. Furthermore, if mixing between $i$ and $j$ occurs on an external line at $\Mm_a$, the transition factor
\begin{equation}
 \Zz_{ij}^{a} = \frac{\Delta_{ij}\ps}{\Delta_{ii}\ps}\bigg\rvert_{\psq=\Mm_a} \label{eq:transij}
\end{equation}
restores the correct normalisation.
The products of the overall normalisation factors $\sqrt{\Zz_a}$ and transition ratios $\Zz_{aj}$ (note the difference between $\Zz_{aj}$ and $\Zb_{aj}$),
\begin{align}
\Zb_{aj} &= \sqrt{\Zz_a}\Zz_{aj}\,,
\label{eq:ZiZij}
\end{align}
form the non-unitary matrix $\Zb$.
The imaginary parts of the self-energies of unstable particles, 
evaluated at non-vanishing incoming momentum,
cause $\Zb$ to be non-unitary.
Accordingly, it does not provide a unitary transformation between the lowest order states $h,H,A$
and the loop-corrected mass eigenstates $h_1, h_2, h_3$.

The $\Zb$-matrix allows one
to express the one-particle irreducible (1PI)
vertex functions $\gh {h_a}$ involving an external loop-corrected
mass eigenstate $h_1,h_2,h_3$ as a linear
combination of the 1PI vertex functions of the lowest-order states, $\gh i$:
\begin{align}
 \gh{h_a}&=\Zb_{ah}\gh h+\Zb_{aH}\gh H+\Zb_{aA}\gh A + ...
 \equiv\sqrt{\Zz_a}\,\left(\Zz_{ah}\gh h+\Zz_{aH}\gh H+\Zz_{aA}\gh A\right) + \dots \label{eq:Vertexha},
\end{align}
where the ellipsis denotes additional contributions from the mixing of 
Higgs bosons with Goldstone and vector bosons, which are not contained the $\Zb$-matrix and need to be calculated explicitly if needed.
As in the full propagator matrix $\dmat_{hHA}\ps$, if $\cp$ is conserved in
the Higgs sector, the $\Zb$-matrix is reduced to the $2\times2$ mixing among
the $\cp$ even states $h$ and $H$, with $\Zb_{3A}=1$, $\Zb_{a3}=\Zb_{3i}=0$
for $a=1,2$ and $i=h,H$. If, on the contrary, complex parameters are
present,  all elements of $\Zb$ can have non-vanishing values and represent
the $3\times3$ mixing.

\paragraph{Mixing propagators in terms of the mass eigenstates}
While the propagators $\Delta_{ij}\ps$ contain the full momentum dependence of the lowest-order propagators and the self-energies, the mixing properties and the leading momentum dependence around the complex pole is adequately approximated by the sum of Breit-Wigner propagators of each mass eigenstate in combination with the on-shell wave-function normalisation factors, $\Zb$. The expansion of the full propagator $\Delta_{ij}\ps$ around all of its complex poles 
$\Mm_a=M_{h_a}^{2}-iM_{h_a}\Gamma_{h_a}$
results in\,\cite{Fuchs:2016swt} 
\begin{align}
 \Delta_{ij}(\psq) &\simeq \sum_{a=1}^{3}\Zb_{ai}\,\BW_a(\psq)\,\Zb_{aj} 
\label{eq:ijBWsum}\,,
\end{align}
where the resonant contribution
of a mass eigenstate $h_a,~a=1,2,3$, with the loop-corrected mass $M_{h_a}$ and the total width $\Gamma_{h_a}$ is described by a Breit--Wigner propagator with a constant width,
\begin{equation}
\BW_a\ps:=\frac{i}{\psq - \Mm_{a}} = \frac{i}{\psq - M_{h_a}^2 + i M_{h_a}\Gamma_{h_a}}  \label{eq:BWdef} \,.
\end{equation}
Hence, although the $\Zb$ matrix has been introduced for the correct normalisation of \textit{external} Higgs bosons, it is also applicable to \textit{internal} Higgs bosons to account for the higher-order mixing properties to a high accuracy.
Eq.\,(\ref{eq:ijBWsum}) implies that the amplitude $\mathcal{A}$ of a process with Higgs exchange between the initial $X$ and final state $Y$ can be written as
\begin{align}
 \mathcal{A} &= \sum_{i,j=h,H,A} \gh i^{X} \,\Delta_{ij}\ps\, \gh j^{Y} 
 \simeq\sum_{a=1}^{3}
\gh{h_a}^{X}\,\BW_a\ps\,\gh{h_a}^{Y}\label{eq:VerthaBW} ,
\end{align}
with $\gh{h_a}^{X}=\sum_{i=h,H,A}\Zb_{ai}\gh i^{X}$ as in
Eq.\,(\ref{eq:Vertexha}). The formulation on the RHS of
Eq.\,(\ref{eq:VerthaBW}) allows one
to divide the full amplitude into the separate contributions of each resonance $h_a$:
\begin{align}
 \mathcal{A}_{h_a} \equiv \gh {h_a}^{X}\, \BW_a\ps\, \gh {h_a}^{Y}
 = \sum_{i,j=h,H,A} \gh i^{X}\, \Zb_{ai}\, \BW_a\ps\, \Zb_{aj}\, \gh j^{Y} \label{eq:amplha}\,,
\end{align}
which will be useful in Sect.\,\ref{sect:int} to compare
the coherent and incoherent sums of amplitudes of several mass eigenstates.

\subsection{Resonant mixing-enhancement of cross sections}
\label{sect:mixenhance}

Since the $\Zb$-matrix is non-unitary, cross sections including mixing
contributions (with $\Zb$ from Eqs.\,(\ref{eq:Zha}-\ref{eq:ZiZij})) can be
enhanced compared to the case without mixing, which would correspond to
a diagonal $\Zb$-matrix (with $\Zb=\mathbb{1}$ for the case where
the external states are properly normalised). 
In particular, in the case of quasi-degenerate states a resonance
enhancement is possible as a consequence of significant off-diagonal and
imaginary parts of the self-energies and the $\Zb$-factors.

The $3 \times 3$ mixing occurring in the general case of complex parameters
can change the phenomenology very significantly as compared to the 
$\cp$-conserving case of real parameters. This is in particular due to the
$\cp$-violating mixing of $H,A$ into $h_2, h_3$ 
since the heavy Higgs bosons $h_2, h_3$ are almost
mass-degenerate in the wide range of parameter space characterised by the
decoupling limit. This degeneracy and the properties of $h_2$
and $h_3$ lead to a significant $H,A$-mixing and an enhancement of the
squared amplitudes. 

For a large mass difference between 
the two heavy states $H$ and $A$ and the light state $h$,
the mixing of $h$ with $H$ and $A$ can be neglected in good approximation.
In this limit,
the squared amplitudes 
of the production processes of
$h_2$ and $h_3$,
$\left|\mathcal{P}_{h_a}\right|^2$ for $a=2,3$,
scale with 
$|\Zb_{aH}|^2 \,|\gh H|^2 +|\Zb_{aA}|^2 \,|\gh A|^2$,
which can become significantly larger than $|\gh i|^2$, $i=H,A$.
In the case of $|\gh H| \simeq |\gh A|$, this scaling factor further simplifies to 
$|\Zb_{aH}|^2+|\Zb_{aA}|^2$, which can exceed 1.
Hence, the cross sections of $\sigma(pp\rightarrow h_2)$ and 
$\sigma(pp\rightarrow h_3)$ 
receive an enhancement 
from the mixing contributions.
We are 
particularly interested here in the production modes 
in association with a pair of bottom quarks ($\bb$) and via gluon fusion
($gg$).
\begin{figure}[ht!]
 \begin{center}
  \subfigure[]{\includegraphics[width=0.243\textwidth]{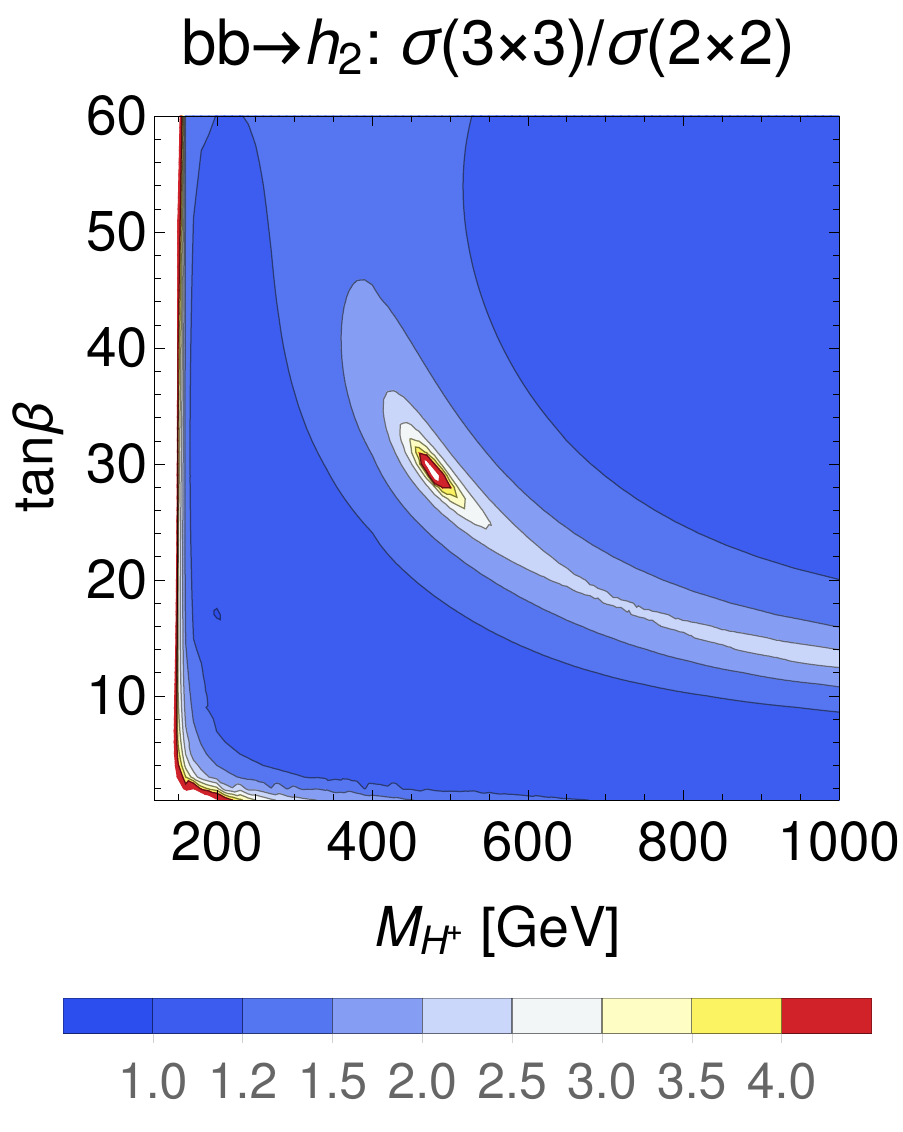}\label{fig:bb2}}
  \subfigure[]{\includegraphics[width=0.243\textwidth]{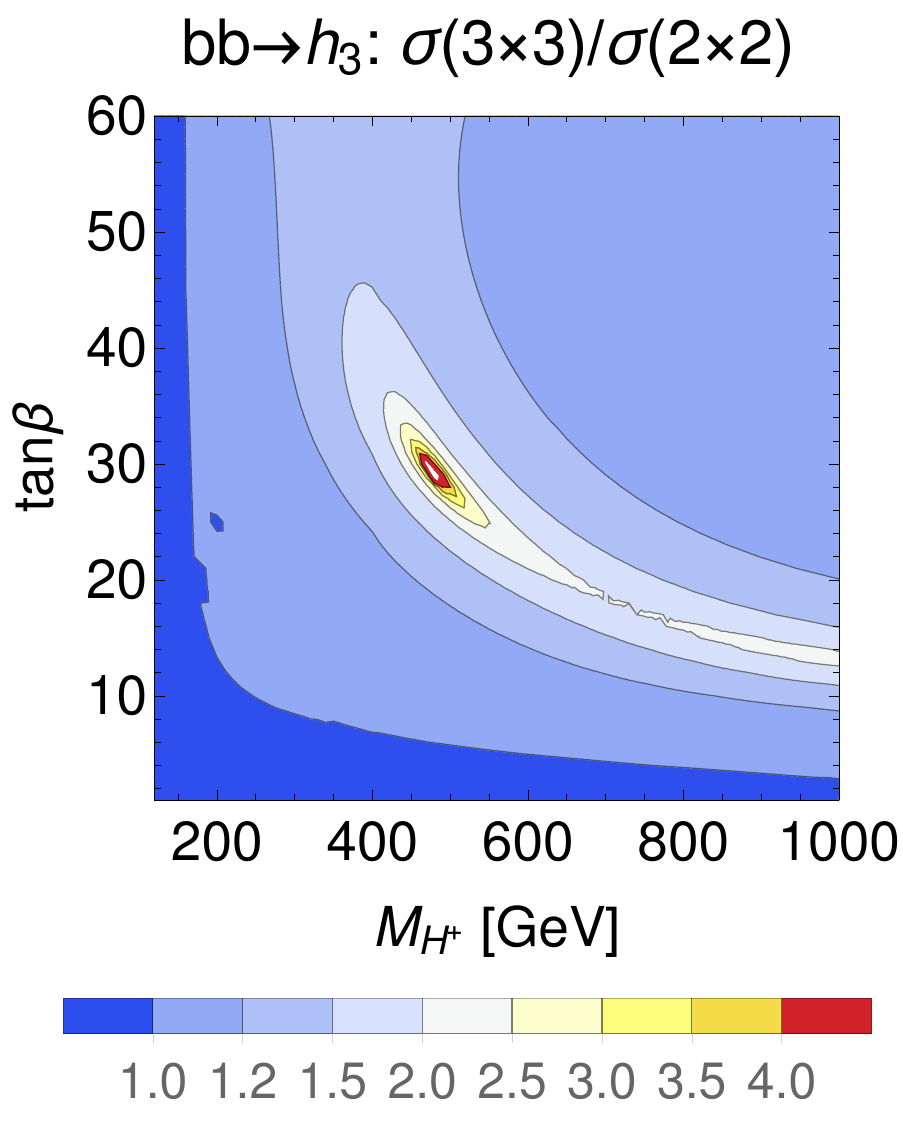}\label{fig:bb3}}
  \subfigure[]{\includegraphics[width=0.243\textwidth]{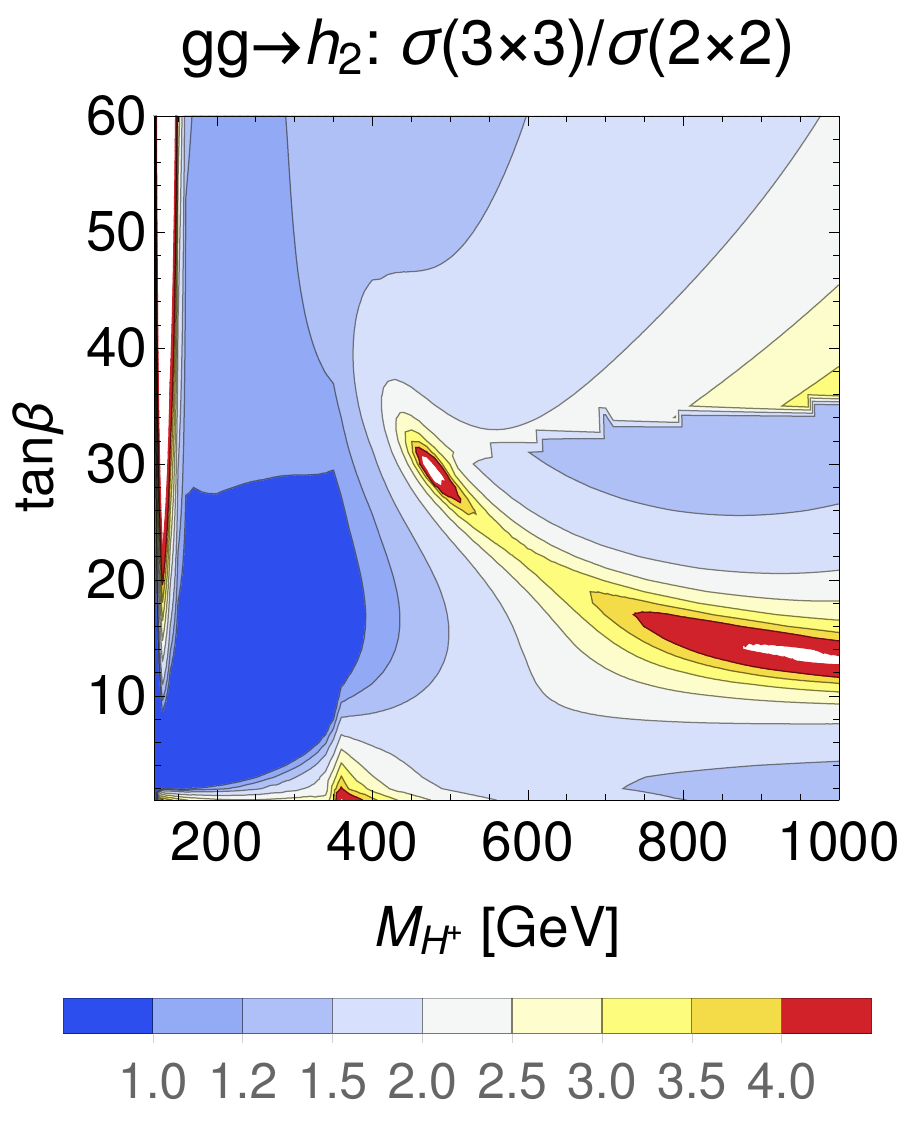}\label{fig:gg2}}
  \subfigure[]{\includegraphics[width=0.243\textwidth]{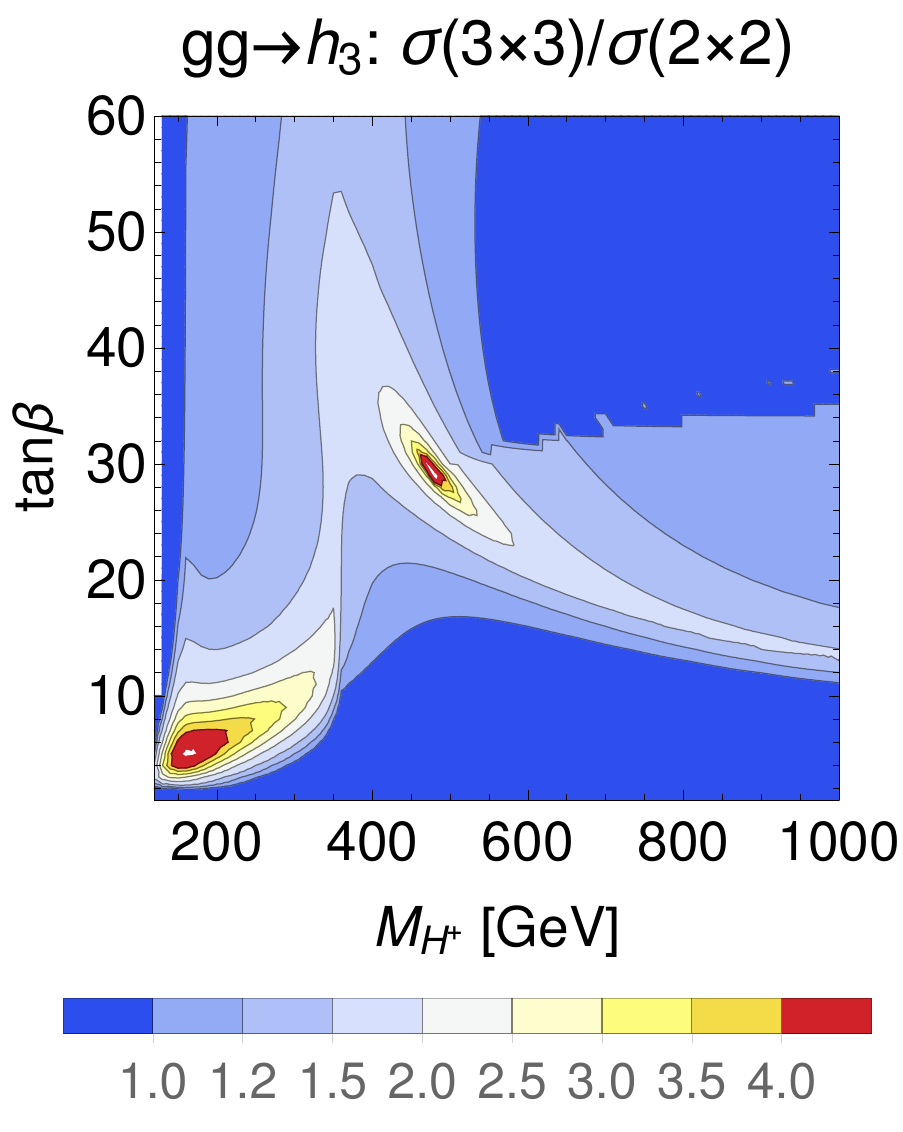}\label{fig:gg3}}
  \caption{Ratios of cross sections $\sigma(P\rightarrow h_a)_{3\times3}/\sigma(P\rightarrow h_a)_{2\times2}$ with $3\times3$ mixing ($\cmhmod$ scenario with $\pat=\pi/4$) 
  in the $\Zb$ factors vs. $2\times2$ mixing ($\pat=0$) for the production modes $P=\bb$ (a,b) and $P=gg$ (c,d) with $h_a=h_2$ (a,c) and $h_3$  (b,d).}
  \label{fig:ratio32}
 \end{center}
\end{figure}

As an example case featuring such an enhancement effect, we illustrate the
$\cp$-violating mixing in a modified version of the $\mhmod$ scenario,
originally defined in Ref.\,\cite{Carena:2013ytb}, where we
introduce as in Refs.\,\cite{Fuchs:2015jwa,Fuchs:2016swt,Fuchs:2016dyn} the complex phase
$\pat=\frac{\pi}{4}$, set $\phi_{A_b}=\phi_{A_{\tau} }=\pat$, 
and increase the value of the higgsino mass parameter
to $\mu=1000\gev$ (as also proposed in Ref.\,\cite{Carena:2013ytb}), which
amplifies the $\cp$-violating terms. 
The input parameters being varied
are $\mhp$ and $\tan\beta$. We refer to this scenario as $\cmhmod$.

Fig.\,\ref{fig:ratio32} shows the single-particle production cross sections
$\sigma(P\rightarrow h_a)$
for $h_2$ and $h_3$ in association with a pair of bottom quarks and via gluon 
fusion. The cross sections for the $\cmhmod$ scenario with $\pat=\pi/4$,
denoted by $3\times3$, have been
normalised in Fig.\,\ref{fig:ratio32} to the ones for the corresponding
$\cp$-conserving scenario with $\pat=0$, 
denoted by $2\times 2$. The predictions for the cross sections have been
obtained with \texttt{FeynHiggs}.
The displayed results for the ratios
$R_{\sigma}(P,h_a):=
\sigma(P\rightarrow h_a)_{3\times3}/\sigma(P\rightarrow h_a)_{2\times2}$
show enhancements by factors of more than 5 in certain areas of
parameter space where $h_2$ and $h_3$ are highly admixed states, especially
around $\mhp\simeq 480\gev$ and $\tanb\simeq 29$. Accordingly, the
enhancement can indeed be related to a resonance-type mixing between the
$\cp$ eigenstates $H$ and $A$.
The observed patterns are similar for
$h_2$ (Figs.\,\ref{fig:bb2}, \ref{fig:gg2}) and 
$h_3$ (Figs.\,\ref{fig:bb3}, \ref{fig:gg3}) 
and, to a lesser extent, also for
production via $\bb$ (Figs.\,\ref{fig:bb2}, \ref{fig:bb3}) and $gg$
(Figs.\,\ref{fig:gg2}, \ref{fig:gg3}). The observed similarities
are related to the
involved pattern of $\Zb$-factors, as discussed above.
The differences between 
the production via $\bb$ and $gg$ arise from the structure of the
amplitudes. While the production in association with a pair of bottom quarks
is dominated by the respective Higgs coupling to bottom quarks (where the
phase $\pat$ enters via the correction to the relation between the
bottom-quark mass and the bottom Yukawa coupling, $\Delta_b$, see e.g.\
Ref.\,\cite{Liebler:2016ceh}), the loop-induced production via gluon
fusion at leading order comprises
contributions from the bottom- and top-quark loops as well
as from their scalar superpartners (the latter induce an explicit 
dependence on $\pat$).%
\footnote{In the $\cp$-conserving case $A$-boson production in gluon fusion 
does not receive a squark-loop contribution at leading-order.}
Those contributions from up-type (s)fermions to gluon fusion
have a sizeable impact 
at lower values of $\tanb$. The edge
along $\tanb\simeq 30$ for $\mhp\gtrsim 600\gev$ in the plots for $gg$-induced
production is caused by the swap of the composition of $h_2$ and $h_3$ being
mostly $H$- or $A$-like, respectively. 

While so far we have restricted our discussion to the individual cross
sections of $h_2$ and $h_3$, in the next section we will analyse the 
$\cp$-violating interference contributions to the full process involving
production and decay.

\section{Interference contributions to the processes
$\left\lbrace \bb, gg \right\rbrace \rightarrow h_1, h_2, h_3 
\rightarrow \tau^+\tau^-$}
\label{sect:int}

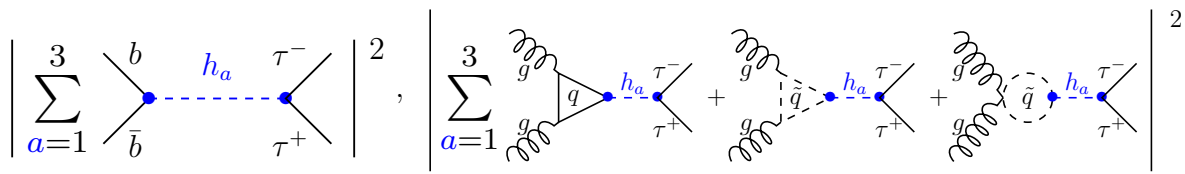
\begin{figure}[h!]
 \begin{center}
 \begin{tikzpicture}[line width=0.7 pt, scale=0.6]
	\draw (-2,1.3)--(-2,-1.3);
	\node[scale=1.5] at (-1,0) {$\sum\limits_{{\color{blue}a}=1}^3$};
	\draw (0,1)--(1,0)--(0,-1);
	\node[color=blue] at (1,0) {\textbullet};
	\node[color=blue] at (4,0) {\textbullet};
	\node at (0.7,1) {$b$};
	\node at (0.7,-1) {$\bar{b}$};
	\draw[color=blue,dashed] (1,0)--(4,0);
	\draw (5,1)--(4,0)--(5,-1);
	\node at (4.1,1) {$\tau^{-}$};
	\node at (4.1,-1) {$\tau^{+}$};
	\node[color=blue] at (2.5,0.7) {$h_a$};
	\draw (5.5,1.3)--(5.5,-1.3);
	\node at (6,1.1) {$2$};
	\node at (6.5,0) {$,$};
\end{tikzpicture}
\begin{tikzpicture}[line width=0.6pt,scale=0.65, every node/.style={scale=0.8}]
\begin{scope}[shift={(-0.8,0)}]
\draw (-0.8,1.3)--(-0.8,-2);
 \node[scale=1.8] at (0,-0.5) {$\sum\limits_{{\color{blue}a}=1}^3$};
\end{scope}
 \draw[gluon] (0,0.8)--(1,0);
 \draw[gluon] (0,-1.8)--(1,-1);
    \draw (2,-0.5)--(1,0)--(1,-1)--(2,-0.5);
    \node at (1.3,-0.5) {$q$};
 \draw[color=blue,dashed] (2,-0.5)--(3,-0.5); 
  \node[color=blue] at (2.5,-0.2) {$h_a$};
  	\node[color=blue] at (2,-0.5) {\textbullet};
	\node[color=blue] at (3,-0.5) {\textbullet};
 \draw (3.7,-1.2)--(3,-0.5)--(3.7,0.2);
 \node at (0.3,0) {$g$};
 \node at (0.3,-1.1) {$g$};
 \node at (3.2, 0.1) {$\tau^{-}$};
 \node at (3.2,-1.1) {$\tau^{+}$}; 
 \begin{scope}[shift={(4.5,0)}]
	\node at (-0.3,-0.5) {$+$};
    \draw[gluon] (0,0.8)--(1,0);
    \draw[gluon] (0,-1.8)--(1,-1);
    \draw[dashed] (2,-0.5)--(1,0)--(1,-1)--(2,-0.5);
	\node at (1.3,-0.5) {$\tilde{q}$};
    \draw[color=blue,dashed] (2,-0.5)--(3,-0.5); 
	\node[color=blue] at (2.5,-0.2) {$h_a$};
  	\node[color=blue] at (2,-0.5) {\textbullet};
	\node[color=blue] at (3,-0.5) {\textbullet};
    \draw (3.7,-1.2)--(3,-0.5)--(3.7,0.2);
      \node at (0.3,0) {$g$};
      \node at (0.3,-1.1) {$g$};
      \node at (3.2, 0.1) {$\tau^{-}$};
      \node at (3.2,-1.1) {$\tau^{+}$}; 
 \end{scope}
  \begin{scope}[shift={(9,0)}]
	\node at (-0.3,-0.5) {$+$};
    \draw[gluon] (0,0.8)--(1,-0.5);
    \draw[gluon] (0,-1.8)--(1,-0.5);
    \draw[dashed] (1.5,-0.5) circle (0.5cm);
	\node at (1.5,-0.5) {$\tilde{q}$};
    \draw[color=blue,dashed] (2,-0.5)--(3,-0.5); 
	\node[color=blue] at (2.5,-0.2) {$h_a$};
  	\node[color=blue] at (2,-0.5) {\textbullet};
	\node[color=blue] at (3,-0.5) {\textbullet};
    \draw (3.7,-1.2)--(3,-0.5)--(3.7,0.2);
      \node at (0.1,0) {$g$};
      \node at (0.1,-1.1) {$g$};
      \node at (3.2, 0.1) {$\tau^{-}$};
      \node at (3.2,-1.1) {$\tau^{+}$}; 
 \end{scope}
 	\draw (13,1.3)--(13,-2);
	\node at (13.5,1.1) {$2$};
\end{tikzpicture}
\caption{Higgs boson production at LO via $\bb$ (left) and $gg$ (right) with decay into $\tau^+\tau^-$. The couplings of the mass eigenstates $h_a=h_1, h_2, h_3$ (blue, dashed) to the initial and final state contain a combination of $\Zb$-factors (denoted by blue circles).}
\label{fig:diagrams}
\end{center}
\end{figure}

For the investigation of a full process of Higgs production and decay 
where the Higgs bosons appear as internal propagators,
we calculate the interference of amplitudes with $s$-channel $h_1, h_2, h_3$
exchange. We consider the production processes 
via $\bb$ and $gg$ and the subsequent
decay into a pair of $\tau$-leptons, as illustrated in
Fig.\,\ref{fig:diagrams},
\begin{eqnarray}
  b\bar{b}	~\rightarrow&h_a~\rightarrow~\tau^{+}\tau^{-}\label{eq:bbh},\\
  gg		~\rightarrow&h_a~\rightarrow~\tau^{+}\tau^{-}\label{eq:ggh}.
\end{eqnarray}
In order to be able to make use of precise predictions for the separate
production\,\cite{Liebler:2016ceh} 
and decay\,\cite{Heinemeyer:1998yj}
processes we focus here on the relative interference
contribution. 
For the phenomenological prediction of the complete processes including
interference, higher-order corrections and quark- and gluon-luminosities at
the LHC, we will combine those relative interference contributions with the
computations of hadronic cross sections and branching ratios from the
existing literature including available higher-order corrections. 
Regarding the determination of the relative interference contributions, we 
restrict ourselves to the computation of the relevant 
$2\to2$ partonic processes at leading order (LO). This means that 
process (\ref{eq:bbh}), which involves $b\bar{b}h_a$-associated production,
is treated at tree-level, while the leading contributions to gluon
fusion entering process (\ref{eq:ggh}) consist of one-loop diagrams with
quarks $q$ and squarks $\tilde q$, in
particular those of the third generation. Besides, we take propagator-type
corrections into account by using Higgs masses, total widths and
$\Zb$-factors from \texttt{FeynHiggs} including the full one-loop and dominant
two-loop corrections. The procedure to evaluate the relative interference
contributions on the basis of leading-order diagrams is motivated by the
fact that the vertex corrections to the production and
decay factorise (apart from non-factorisable initial-to-final state
radiation). It furthermore avoids double-counting of higher-order
contributions that are incorporated in the separate production and decay
processes. We calculate the LO contributions to the
amplitude and the cross section of the full process of production and decay using
\texttt{FeynArts}\,\cite{Kublbeck:1990xc,Denner:1992vza, Kublbeck:1992mt,
Hahn:2000kx, Hahn:2001rv} with a model file containing $\Zb$-factors for the
Higgs vertices (in this way the Higgs propagators are expressed in terms of 
$\Zb$-factors and lowest-order Breit--Wigner propagators as given in
(\ref{eq:ijBWsum})),
\texttt{FormCalc}\,\cite{Hahn:1998yk,
Hahn:1999wr,Hahn:2000jm, Hahn:2006qw, Hahn:2006zy}
and, as mentioned above, \texttt{FeynHiggs} for quantities from the Higgs
sector.

\paragraph{Quantifying the interference}
In order to determine the interference term in each of the two processes (\ref{eq:bbh}, \ref{eq:ggh}), we distinguish between the \textit{coherent} sum of the $2\rightarrow2$ amplitudes with $h_1,h_2,h_3$-exchange including the interference
and their \textit{incoherent sum} without the interference,
\begin{align}
 |\mathcal{A}|^{2}_{\textrm{coh}} =  \bigg\lvert \sum_{a=1}^{3} \mathcal{A}_{h_a} \bigg\rvert^{2} \,,
 ~~~~~~~~~~~
 |\mathcal{A}|^{2}_{\textrm{incoh}} =   \sum_{a=1}^{3} \bigg\lvert\mathcal{A}_{h_a} \bigg\rvert^{2} \label{eq:amplcoh}\,.
 \end{align}
The interference term is then obtained from the difference,
\begin{align}
 |\mathcal{A}|^{2}_{\textrm{int}} = |\mathcal{A}|^{2}_{\textrm{coh}} -|\mathcal{A}|^{2}_{\textrm{incoh}}
  = \sum_{a<b}2\,\textrm{Re}\left[\mathcal{A}_{h_a} \mathcal{A}_{h_b}^{*}\right].
\end{align}
Accordingly, the coherent cross section refers to the cross section based on the coherent sum of amplitudes,
$\sigma_{\textrm{coh}}\equiv \sigma\left(|\mathcal{A}|^{2}_{\textrm{coh}}\right)$, 
and likewise for the incoherent cross section 
$\sigma_{\textrm{incoh}}$ (omitting the interference term) 
and the interference part of the cross section 
$\sigma_{\textrm{int}}$.
Furthermore, we define the relative interference term for each production mode $P=\bb, gg$ as
\begin{eqnarray}
 \eta^P :=\frac{\sigma^P_{\rm{int}}}{\sigma^P_{\rm{incoh}}}\label{eq:eta}\,.
\label{eq:etaPdef}
\end{eqnarray}
Beyond this overall interference term, it is also useful to differentiate
between the three ``squared'' terms of each Higgs boson $h_a$ in
$\sigma_{h_a}$ and the three interference terms $\sigma_{\rm{int}_{ab}}$
involving two Higgs bosons $h_a, h_b$ out of the three, respectively, again
separately for both production modes $P$,
\begin{align}
 \sigma^{P} &=\sigma^{P}_{h_1}+\sigma^{P}_{h_2}+\sigma^{P}_{h_3}
 +\sigma^{P}_{\rm{int}_{12}} +\sigma^{P}_{\rm{int}_{13}} +\sigma^{P}_{\rm{int}_{23}}\label{eq:sigmaP}.
\end{align}
In order to express the cross section into separate Higgs contributions (as
in the incoherent case), we introduce relative interference factors
$\eta_a^P$ that rescale the individual Higgs rates:
\begin{align}
 \sigma^{P} &=\sigma^{P}_{h_1}\,(1+\eta_1^P) +
\sigma^{P}_{h_2}\,(1+\eta_2^P) + \sigma^{P}_{h_3}\,(1+\eta_3^P)\,,
\label{eq:rescaled}
\end{align}
where we define $\eta_a^P$ by weighting $\sigma_{\rm{int}_{ab} }^{P}$ according to the $\sigma_{h_a},~\sigma_{h_b}$ contributions with the weight (for $a,b,c=1,2,3$)
\begin{align}
 w_{ab}^{P,a} = \frac{\sigma_{h_a }^{P}}{\sigma^{P}_{h_a}+\sigma^{P}_{h_b}} \equiv 1 -  w_{ab}^{P,b}\,,
\end{align}
resulting in
\begin{align}
 \eta^{P}_a &= \frac{w_{ab}^{P,a}\,\sigma^{P}_{\rm{int}_{ab}} + w_{ac}^{P,a}\,\sigma^{P}_{\rm{int}_{ac}}}{\sigma_{h_a}^{P}}
 = \frac{\sigma^{P}_{\rm{int}_{ab}}}{\sigma_{h_a}^{P}+\sigma_{h_b}^{P}}
   +\frac{\sigma^{P}_{\rm{int}_{ac}}}{\sigma_{h_a}^{P}+\sigma_{h_c}^{P}}\label{eq:etaP}\,.
\end{align}
This definition of $\eta_a^P$ is stable also in the case that one of the individual
contributions $\sigma_{h_a}^P$ is suppressed.

The relative interference contributions $\eta_a^P$ 
can then be applied to obtain a result consisting of the separate
predictions for the production and decay processes and the respective
interference contributions
\begin{equation}
 \sigma(pp\rightarrow P\rightarrow h_{1,2,3}\rightarrow \tau^+\tau^-) \simeq 
 \sum_{a=1}^3 \sigma(pp\rightarrow P\rightarrow h_a)\cdot (1+\eta_a^P)\cdot
\br(h_a\rightarrow \tau^+\tau^-)\,.
\label{eq:relint1}
\end{equation}

In order to confront the theoretical predictions in the considered scenario
with the existing experimental bounds we use the
program \texttt{HiggsBounds-4.2.0}\,\cite{\HBlabel}. For the incorporation of the
interference contributions it is convenient to
rescale the ratio of the production cross sections $P\rightarrow h_a$ in 
the MSSM with respect to the SM,
which is supplied as an input to \texttt{HiggsBounds}, in the following way,
\begin{eqnarray}
 \frac{\sigma^{\rm{MSSM}}(P\rightarrow h_a)}{\sigma^{\rm{SM}}(P\rightarrow h_a)}
~~ \longrightarrow ~~
 \frac{\sigma^{\rm{MSSM}}(P\rightarrow h_a)}{\sigma^{\rm{SM}}(P\rightarrow
h_a)}\,\cdot (1+\eta_a^P)\,.
\label{eq:relint2}
\end{eqnarray}
The choice of associating the interference contributions with the production
processes in \texttt{HiggsBounds} while leaving the 
branching ratios $\textrm{BR}(h_a\rightarrow \tau^{+}\tau^{-})$ unchanged 
is of course of purely technical nature. One always needs to ensure that
interference contributions are properly taken into account for the full
process including production and decay, see Ref.~\cite{IntCalc:InProgress}.

\begin{figure}[ht!] 
 \begin{center}
  \subfigure[]{\includegraphics[width=0.48\textwidth]{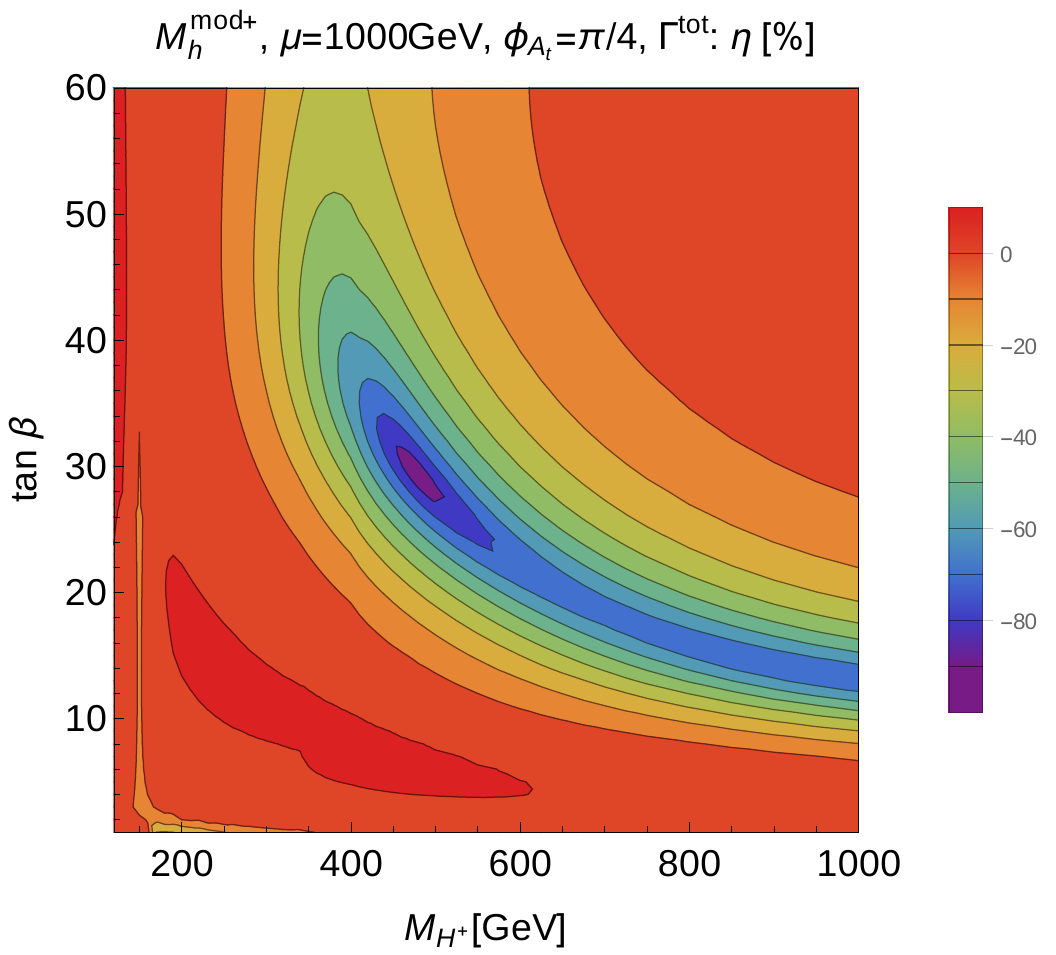}}
  \subfigure[]{\includegraphics[width=0.48\textwidth]{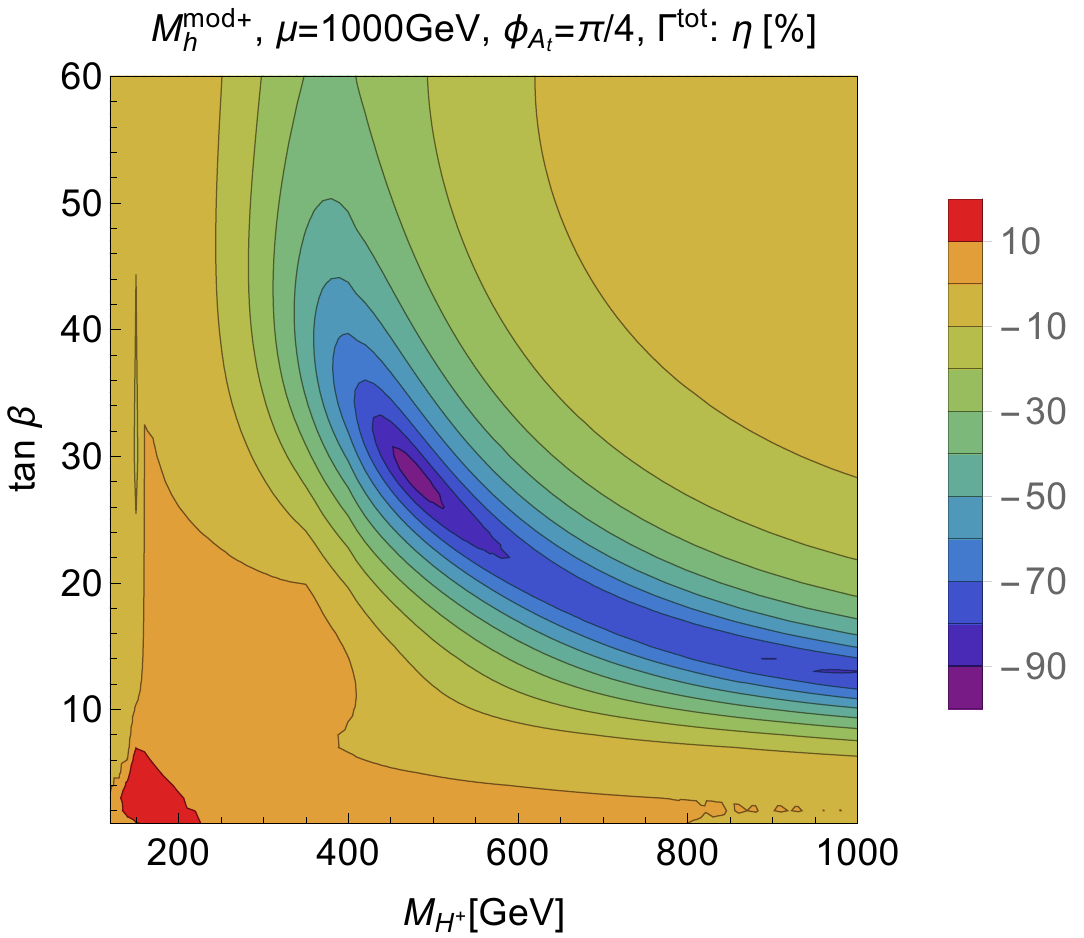}}
  \caption{Relative interference contribution $\eta[\%]$ of the Higgs bosons $h_1,h_2,h_3$ decaying to $\tau^{+}\tau^{-}$ in the complex $\mhmod$ scenario with $\mu=1000\gev$ and $\pat=\frac{\pi}{4}$. (a) $b\bar{b}$ initial state, (b) $gg$ initial state (note the different scale of the colour code).}
 \label{fig:eta}
 \end{center}
\end{figure}

For the $\cmhmod$ scenario the results $\eta^{\bb}$ and $\eta^{gg}$ 
as defined in (\ref{eq:etaPdef}), given in $\%$,
are shown in Fig.\,\ref{fig:eta} within the ($\mhp$, $\tanb$) parameter planes.
The
interference effect in both processes exhibits a similar pattern. First of
all, it is \textit{destructive}, i.e.\ $\eta^P<0$, throughout the parameter
plane (apart from the small dark red regions). Due to the approximate mass
degeneracy of $h_2$ and $h_3$ and the sizeable $H-A$ mixing, the
interference contribution becomes in both processes very large in significant parts
of the parameter space. At $\mhp\simeq 480\gev$ and $\tanb\simeq29$, the
relative interference contribution reaches a minimum of $\eta^P\simeq-97\%$
in both processes
so that the cross section is almost completely cancelled by the drastic,
negative interference term. This strongest interference region is surrounded
by a ``valley'' of also substantial destructive interference.

\paragraph{Conditions for a sizeable interference term}
\begin{figure}[ht!]
 \begin{center}
  \subfigure[]{\includegraphics[width=0.495\textwidth]{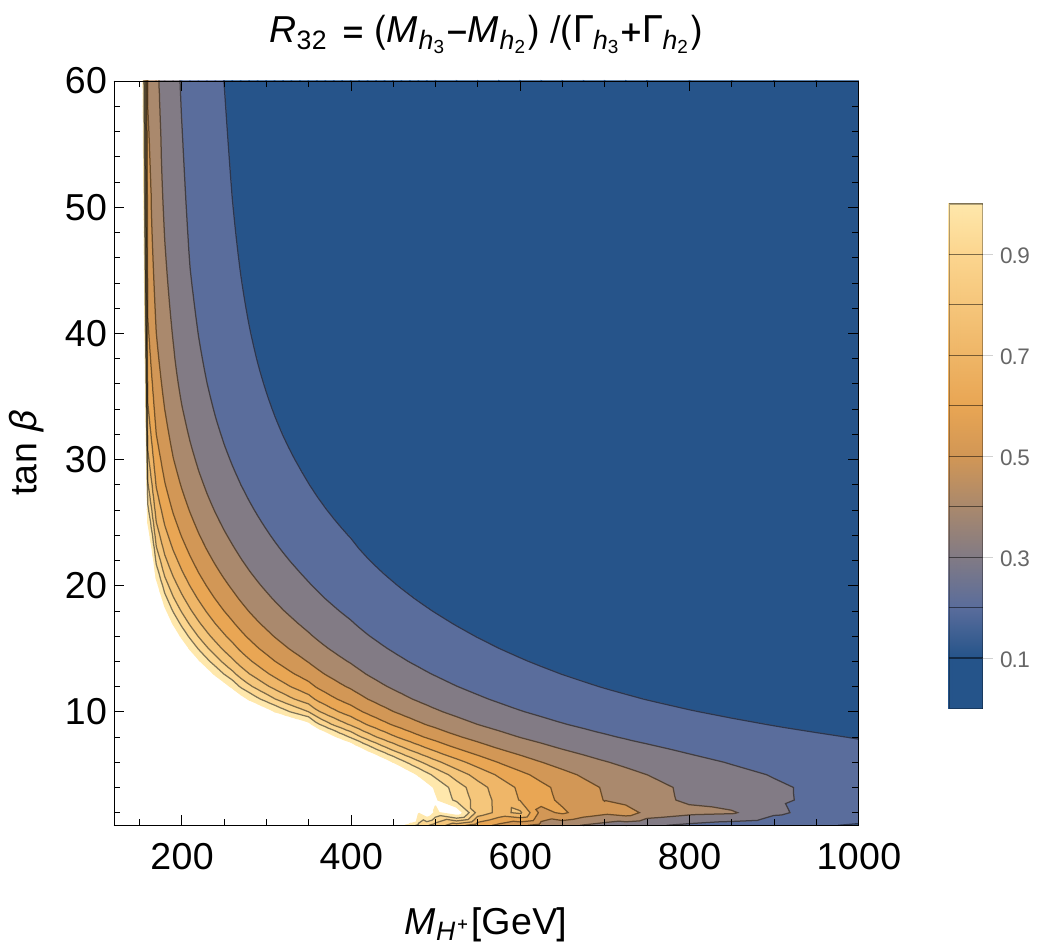}\label{fig:R32}}
  \subfigure[]{\includegraphics[width=0.495\textwidth]{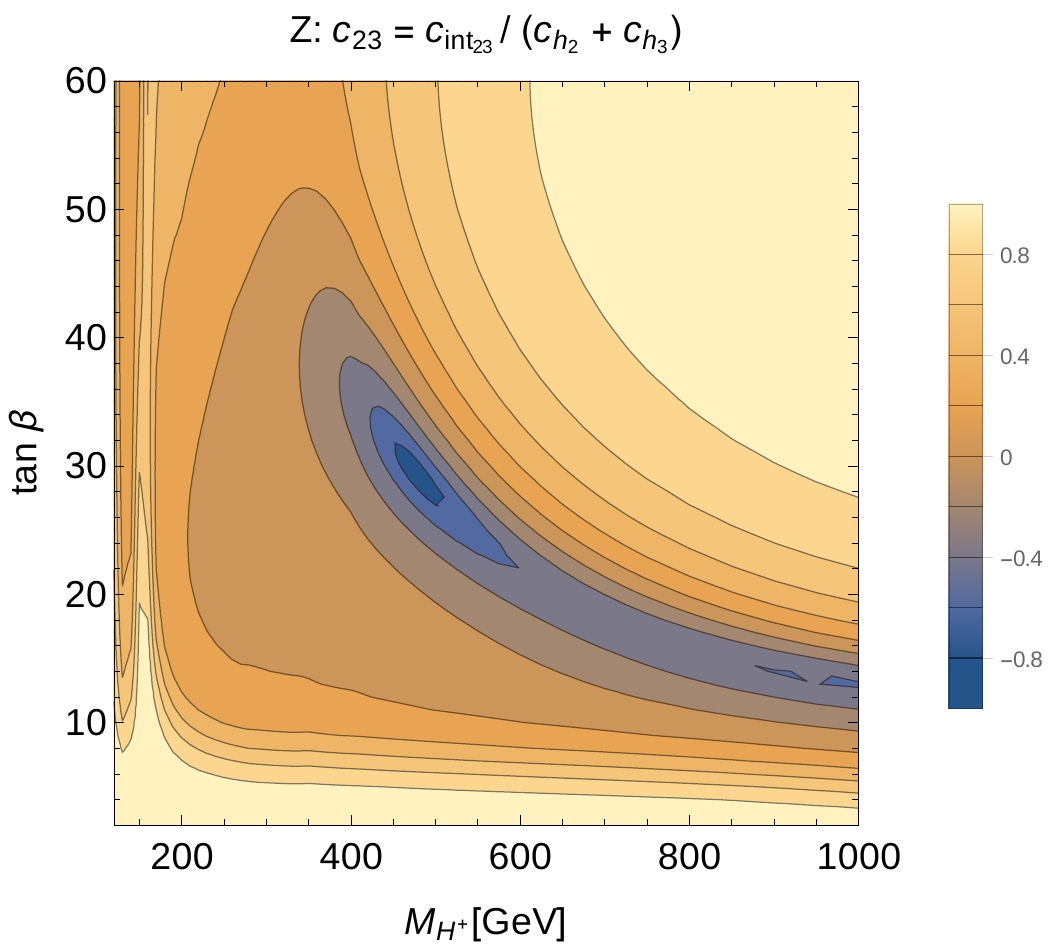}\label{fig:xZ}}\\
  \caption{\textbf{(a):} Ratio $R_{32}$ of mass difference $M_{h_3}-M_{h_2}$ and sum of total widths $\Gamma_{h_2}+\Gamma_{h_3}$.
  \textbf{(b):} Ratio $c_{23}$ of couplings in the interference term of the
$\bb$-initiated process 
compared to those in the incoherent sum, including
$\Zb$-factors.}
 \end{center}
\end{figure}
Having observed the occurrence of a large destructive interference
contribution at the example of the $\cmhmod$ scenario in
Fig.\,\ref{fig:eta}, we now investigate in more detail
under which conditions such an interference can arise.
An important criterion is the overlap of two resonances which is determined
by the mass splitting and the total widths. In the decoupling limit, $h_2$
and $h_3$, which are involved in the relevant interference, are quasi
degenerate while $h_1$ is much lighter. Because of the significant total widths
$\Gamma_{h_2}, \Gamma_{h_3}$, the $h_2-h_3$ overlap 
\begin{equation}
R_{32}:=\frac{M_{h_3}-M_{h_2}}{\Gamma_{h_2}+\Gamma_{h_3}}< 1 
\end{equation}
is also fulfilled for not exactly degenerate states as displayed in
Fig.\,\ref{fig:R32}.

In order to understand the location of the strongest interference, we examine the couplings that play a role in the interference term compared to those in the incoherent sum of the $\bb$-initiated process:
\begin{align}
 c_{23} = \frac{2\textrm{Re}[g_{h_2\tau\tau}\,g_{h_2 bb}\, g^{*}_{h_3\tau\tau}\,g^{*}_{h_3 bb}]}{|g_{h_2\tau\tau}\,g_{h_2 bb}|^{2} + |g_{h_3\tau\tau}\,g_{h_3 bb}|^{2}} \label{eq:c23ratio},
\end{align}
where $g_{h_af\bar f}$ with $a=1,2,3,~f=\tau,b$ are the tree-level couplings $g_{if\bar f}$ from Eq.\,(\ref{eq:GHtt}) for $i=h,H,A$, combined with two-loop $\Zb$-factors from \texttt{FeynHiggs} according to Eq.\,(\ref{eq:Vertexha}):
\begin{align}
 g_{h_a f\bar f} = \sum_{i=h,H,A} \Zb_{ai} g_{i f\bar f}.
\end{align}
Since the masses $m_{\tau}, m_b$ and other constants cancel out in
Eq.\,(\ref{eq:c23ratio}), the ratio $c_{23}$ is determined by the on-shell
$\Zb$-factors and the angles $\cos\alpha$ and $\sin\beta$.
Fig.\,\ref{fig:xZ} shows that the behaviour of $c_{23}$ 
largely determines the pattern of the relative interference contribution
observed in Fig.\,\ref{fig:eta},
whereas effective couplings based on a $\psq=0$ approximation of the
self-energies  
or the pure tree-level couplings 
would yield a completely different pattern, see Ref.\,\cite{Fuchs:2015jwa}.
The interference contribution on the squared matrix element level is
proportional to\,\cite{Fowler:2010eba}
\begin{align}
   |\mathcal{A}|^{2}_{\rm{int}} \propto
   \cb^{-4}\,
  2\textrm{Re}\left[
 (c_{\alpha}^{2}\Zb_{2H}\Zb_{3H}^{*}
 +s_{\beta}^{2}\Zb_{2A}\Zb_{3A}^{*}
 )^{2} \BW_2(s) \BWc_3(s)
 \right]\label{eq:intZZ}.
\end{align}
In the decoupling region of $M_H^\pm\gg M_Z$,
the heavy Higgs bosons $h_2$ and $h_3$ have very similar masses $M_{h_2}\simeq M_{h_3}$ and widths
$\Gamma_{h_2}\simeq\Gamma_{h_3}$ so that the product
$\BW_2(s) \BWc_3(s) \simeq |\BW_2|^{2}$ becomes approximately real. 
In this limit, the relations $\Zb_{2H}\simeq\Zb_{3A}$, $\Zb_{2A}\simeq -\Zb_{3H}$ and $\cos\alpha \simeq \sin\beta$ simplify Eq.\,(\ref{eq:intZZ}) to:
\begin{align}
 |\mathcal{A}|^{2}_{\textrm{int}} \propto
 -8\tb^{4}\, (\textrm{Im}\Zb_{2H}\,\textrm{Re}\Zb_{2A}\,-\, \textrm{Re}\Zb_{2H}\,\textrm{Im}\Zb_{2A})^{2}\,|\BW_2(s)|^{2}
 \label{eq:intZZdecoupling}.
\end{align}
As expected, the interference term of $h_2$ and $h_3$ vanishes for the case
of $\cp$ conservation in the Higgs sector, in which case
$\Zb_{2A}=\Zb_{3H}=0$ holds.
In addition to the $\cp$-violating mixing,
a non-zero interference term requires non-vanishing imaginary parts of the
$\Zb$-factors, which originate from the imaginary parts of the Higgs
self-energies. Consequently, replacing the on-shell $\Zb$-factors by real
mixing factors $\Ub$ (see e.g.\ Refs.\,\cite{Frank:2006yh,Fuchs:2016swt}) in
an effective coupling approach with $\psq=0$ renders the interference term
zero in the decoupling limit even though the $\Ub$-matrix may contain
equally large diagonal and off-diagonal elements.
Even if the conditions of $\cp$-violating mixing and the presence of
non-vanishing imaginary parts are fulfilled, there might still be a
cancellation between the two terms within the bracket in
Eq.\,(\ref{eq:intZZdecoupling}).

Outside the decoupling limit, with unequal masses, widths and mixing properties of $h_2$ and $h_3$, the full product of angles from the couplings, $\Zb$-factors and complex Breit-Wigner functions has to be taken into account. 
However, in the relevant part of the considered parameter plane, the decoupling limit is reached. Given the quasi-degeneracy of $M_{h_2}$ and $M_{h_3}$ shown in Fig.\,\ref{fig:R32}, the structure of the $\Zb$-matrix provides in fact a well-suited indication of the relevance of the interference term. In particular, the square of the bracket and the absolute square of the Breit-Wigner function in combination with the overall minus sign in Eq.\,(\ref{eq:intZZdecoupling}) explain the observed \textit{destructive} interference effect.

\section{Impact of interference and mixing on exclusion bounds}
\label{sect:TheoExp}
As two consequences of the $\cp$-violating mixing, in the previous sections we have investigated the
enhancement of production cross sections for the mass eigenstates 
and the reduction of the complete processes of production and decay
due to destructive interference.
In this section we will compare these theory predictions with experimental
limits from searches for heavy additional Higgs bosons in the $\tau^+\tau^-$
final state at Run~1 of the LHC\,\cite{Aad:2014vgg,CMS:2015mca}. 
In the LHC searches for additional Higgs bosons in supersymmetric models 
so far it has been assumed that $\cp$ is conserved, so that no mixing occurs
between $\cp$-even and $\cp$-odd states. Under this assumption, the
contributions from $\cp$-even and $\cp$-odd states can be added as an
incoherent sum. The question arises in this context in how far the exclusion
bounds that have been obtained under the assumption of $\cp$ conservation
will be modified in the general case where the possibility of
$\cp$-violating mixing and interference effects is taken into account. 
In the following we will first as an illustration compare 
single cross section limits with the theory predictions, before performing a
more detailed analysis of the exclusion limits in the $(\mhp,\tanb)$
parameter plane of the $\cmhmod$ scenario using the tool \texttt{HiggsBounds}.

\subsection{Coherent and incoherent cross sections}
\label{sect:InCoh}
\begin{figure}[ht!]
 \begin{center}
  \subfigure[]{\includegraphics[width=0.495\textwidth]{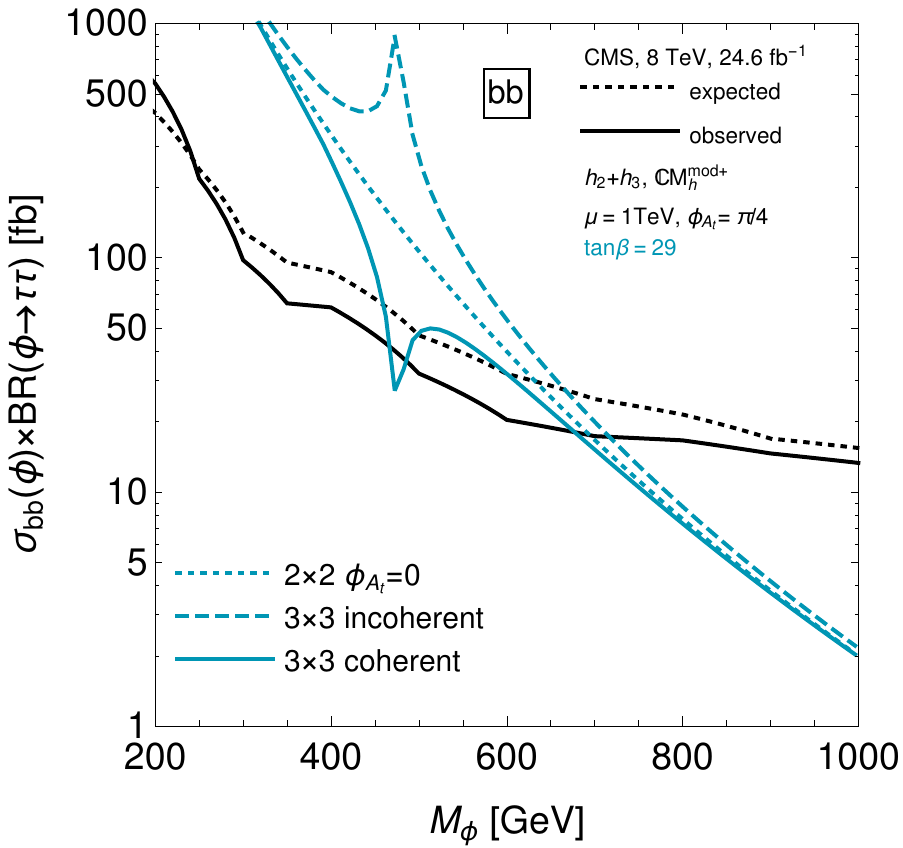}\label{fig:a}}
  \subfigure[]{\includegraphics[width=0.495\textwidth]{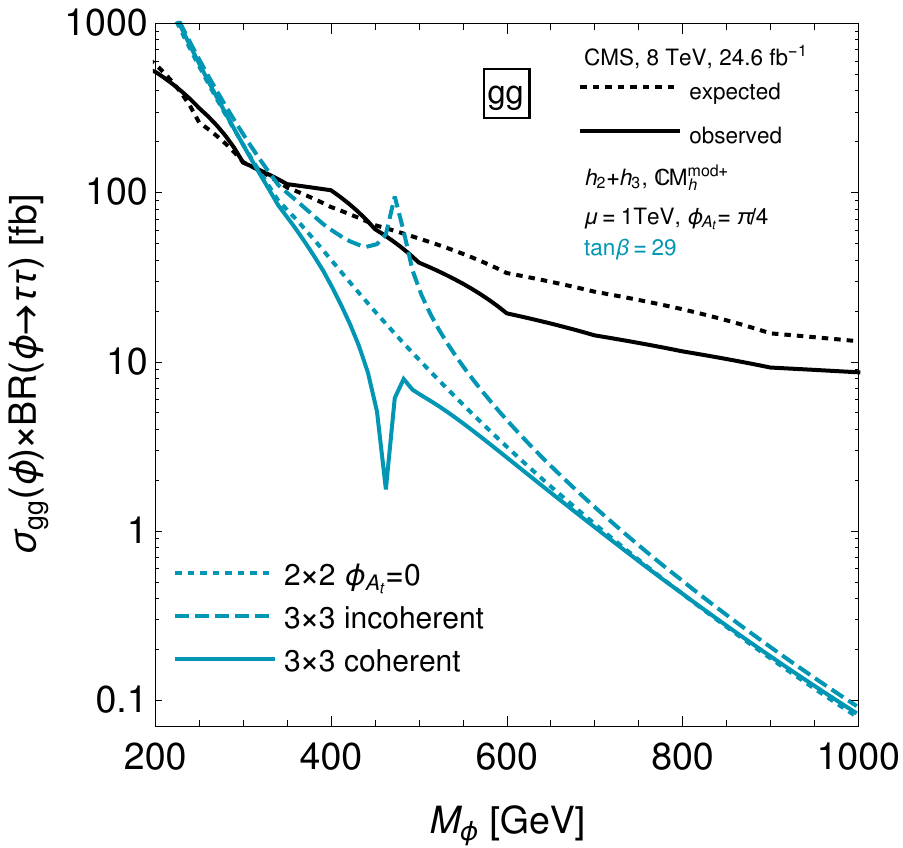}\label{fig:b}}\\
  \subfigure[]{\includegraphics[width=0.495\textwidth]{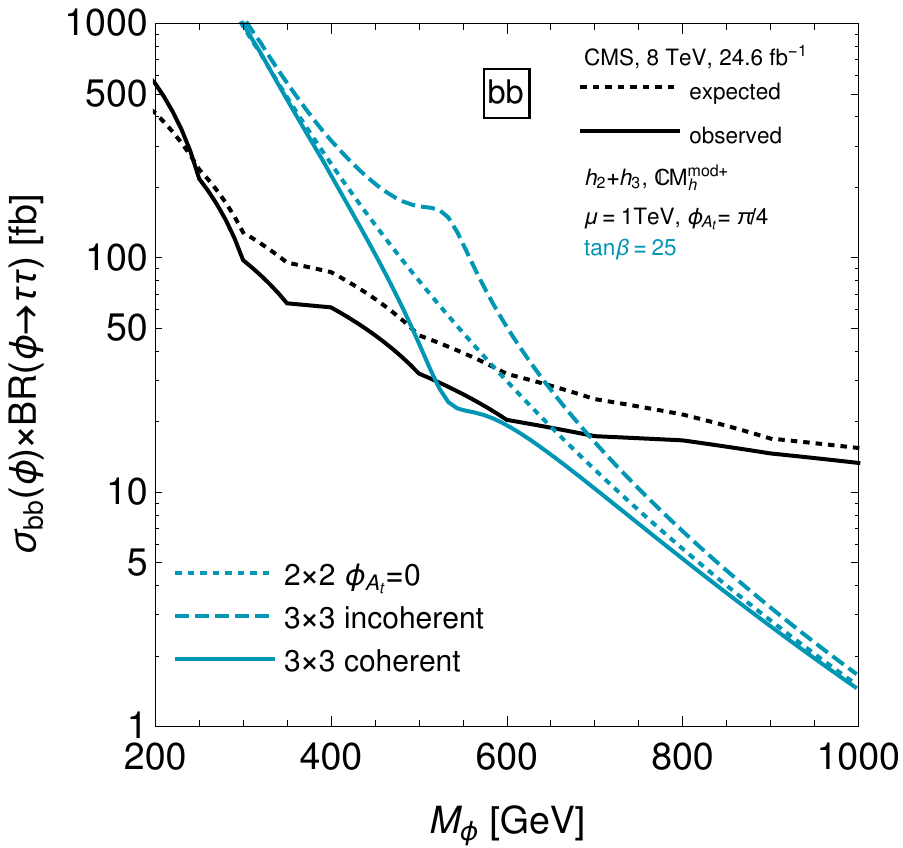}\label{fig:c}}
  \subfigure[]{\includegraphics[width=0.495\textwidth]{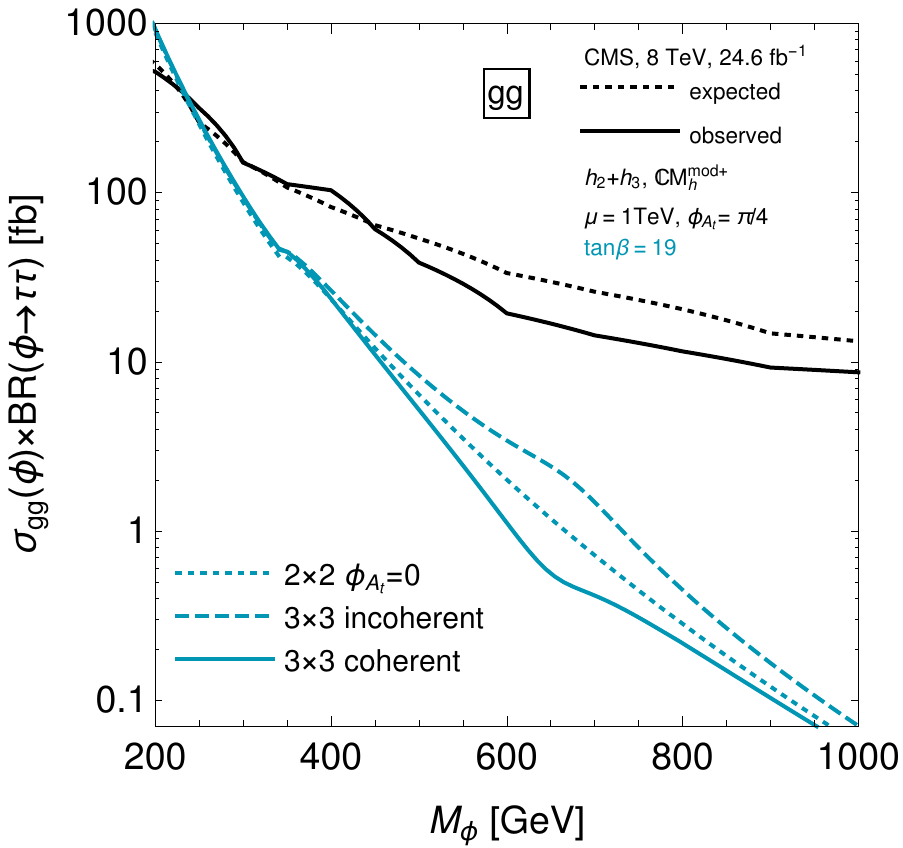}\label{fig:d}}
  \caption{Comparison of predicted Higgs cross sections times branching
ratio into $\tau^+\tau^-$ with and without the interference and with
$2\times 2$ and $3\times 3$ mixing, for a fixed value of $\tan\beta$, to
experimental exclusion bounds. Left column (a, c): production via $\bb$.
Right column (b, d): production via $gg$. Upper row (a, b): strongest
interference effect at $\tb=29$. Lower row: (c) $\tb=25$, (d) $\tb=19$. Each
plot shows the CMS observed (black, solid) and expected (black, dotted)
exclusion bounds at $95\%$ CL at $8\tev$ with $\int\mathcal{L}=24.6\fb^{-1}$
from Ref.\,\cite{CMS-PAS-HIG-13-021,CMS:2015mca,Khachatryan:2014wca}, as
well as the theory prediction in the $\mathds{C}\mhmod$ scenario for the
combined cross section of $h_2$ and $h_3$ 
  as the incoherent sum restricted to $\cp$-conserving mixing for $\pat=0$
(turquoise, dotted, labelled as ``$2\times 2$''), 
  the incoherent sum with $\cp$-violating mixing for $\pat=\pi/4$
(turquoise, dashed, labelled as ``$3\times 3$ incoherent'')
  and the coherent sum (\textit{i.\,e.}\ including the interference term)
with $\cp$-violating mixing for $\pat=\pi/4$
(turquoise, solid, labelled as ``$3\times 3$
coherent'').}
  \label{fig:InCohMphi}
 \end{center}
\end{figure}

In Fig.\,\ref{fig:InCohMphi} we show the theory predictions (turquoise) for
$\sigma(pp\rightarrow P \rightarrow h_2, h_3 \rightarrow \tpm)$
depending on the mass $\mphi$ of a neutral scalar resonance $\phi$ for each
of the production modes $P=\bb,gg$ in comparison with the respective
experimental limits on the production of a single resonance 
(black). The displayed experimental results 
represent the expected (dotted) and observed (solid) CMS exclusion bound at
the $95\%$\,CL from Run~1 of the LHC with $\int\mathcal{L}=24.6\fb^{-1}$ reported in
Ref.\,\cite{CMS-PAS-HIG-13-021} (which is also used in \texttt{HiggsBounds}).

The theory prediction has been calculated using $\mhp$ and $\tanb$ as input
parameters and is shown as a function of $\mphi:=M_{h_3}$ ($\simeq
M_{h_2}$ in the relevant region) for representative values of $\tanb$. The
differences between the incoherent sum restricted to $\cp$-conserving
$2\times2$ mixing for $\pat=0$ 
(dotted), the incoherent sum with $\cp$-violating $3\times3$ mixing for
$\pat=\pi/4$ 
(dashed) and the coherent sum including $3\times3$ mixing and the interference term
for $\pat=\pi/4$ (solid) are visible. 
Figs.\,\ref{fig:a} and \ref{fig:c} show the $bb$-initiated process, whereas
\ref{fig:b} and \ref{fig:d} show the $gg$-initiated one. The two upper plots
correspond to $\tanb=29$, where the strongest interference is observed, while
the lower two plots represent $\tanb=25$ and $\tanb=19$, respectively, with
still a large interference albeit less drastic effect in the $gg$-process.
The comparison of the predicted cross sections times branching ratios with
and without the interference term
to the experimental bound reveals that 
the presence of a non-vanishing value of the phase of the parameter $A_t$ 
in this scenario has a large impact on the exclusion bounds. 
In comparison with the case where $\cp$ conservation is assumed ($\pat=0$), 
the cross section in the case of $3\times 3$ mixing would be significantly
enhanced in the resonance region if only the incoherent sum of the
contributions were taken into account. 
In contrast, the incorporation of the interference
contribution in fact has the overall effect of a reduction of the cross
section in the resonance region. 

In the comparison of the theory predictions with a single cross section
limit considered here,
for a given value of $\tb$ one obtains an exclusion 
at the $95\%$CL for $\mphi$ values where
the prediction of $\sigma\times\br$ is larger than the observed limit.
Accordingly, the exclusion based on the mixing-enhanced $3\times 3$ incoherent 
cross section would be stronger than for the case where $\cp$ conservation is 
assumed ($2\times 2$ with $\pat=0$), while the inclusion of the interference
contribution in the full result leads to weaker bounds than 
for the case where $\cp$ conservation is assumed. Therefore in general the
lower bound on $\mphi$ for a given value of $\tb$ that can be set by 
this comparison is reduced as compared to the case where $\cp$ conservation 
is assumed. 

While this shift of the lower bound on $\mphi$ is only moderate if the
intersection between the prediction and the observed limit occurs outside of
the resonance region, as it is the case 
in Fig.\,\ref{fig:b} and in Fig.\,\ref{fig:d} for the $gg$ production mode
(which is less constraining than the $\bb$ process in the parameter regions
considered here),
the situation is different if
the large destructive interference contribution in the resonance region 
reduces the cross section such that it falls below the observed limit. This
can be seen in Fig.\,\ref{fig:a} for the $\bb$ production mode. In addition
to the shift in the lower bound on $\mphi$ from about 695~GeV for the case 
$\pat=0$ to about 675\,GeV
for the case $\pat=\pi/4$,
in this case a range of unexcluded values of $\mphi$ opens up 
between 465~GeV and 485~GeV
which would appear to be excluded under the assumption of $\cp$ conservation.
A similar effect can be seen in 
Fig.\,\ref{fig:c} for the $\bb$ production mode
for the smaller
value of $\tb$ which is somewhat off the resonance.
As a consequence,
the resulting interference effect and also the enhancement of the
incoherent cross section are less sharp. 
In this case the lower bound on 
$\mphi$ is shifted from about 650~GeV for the case 
$\pat=0$ to about 520~GeV for the case $\pat=\pi/4$.
If instead the interference contribution were neglected for the $3 \times 3$
case, the incoherent cross section would indicate an exclusion on $\mphi$ up
to about 690~GeV in this case.

\begin{figure}[t]
 \begin{center}
  \subfigure[]{\includegraphics[width=0.495\textwidth]{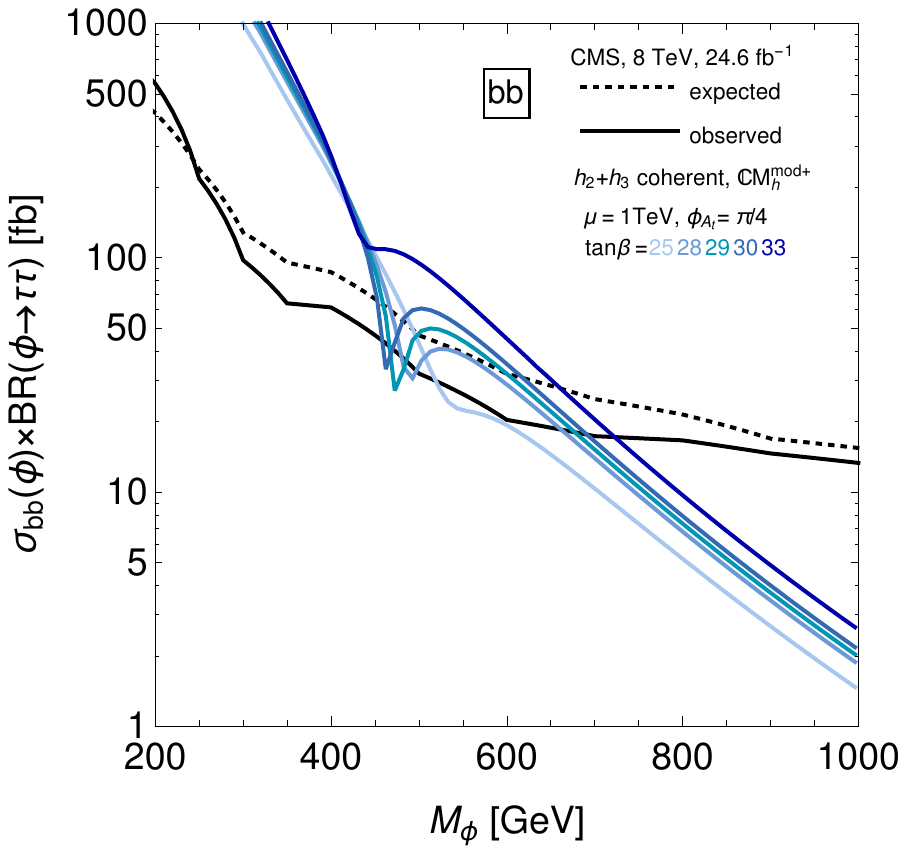}\label{fig:bbtb}}
  \subfigure[]{\includegraphics[width=0.495\textwidth]{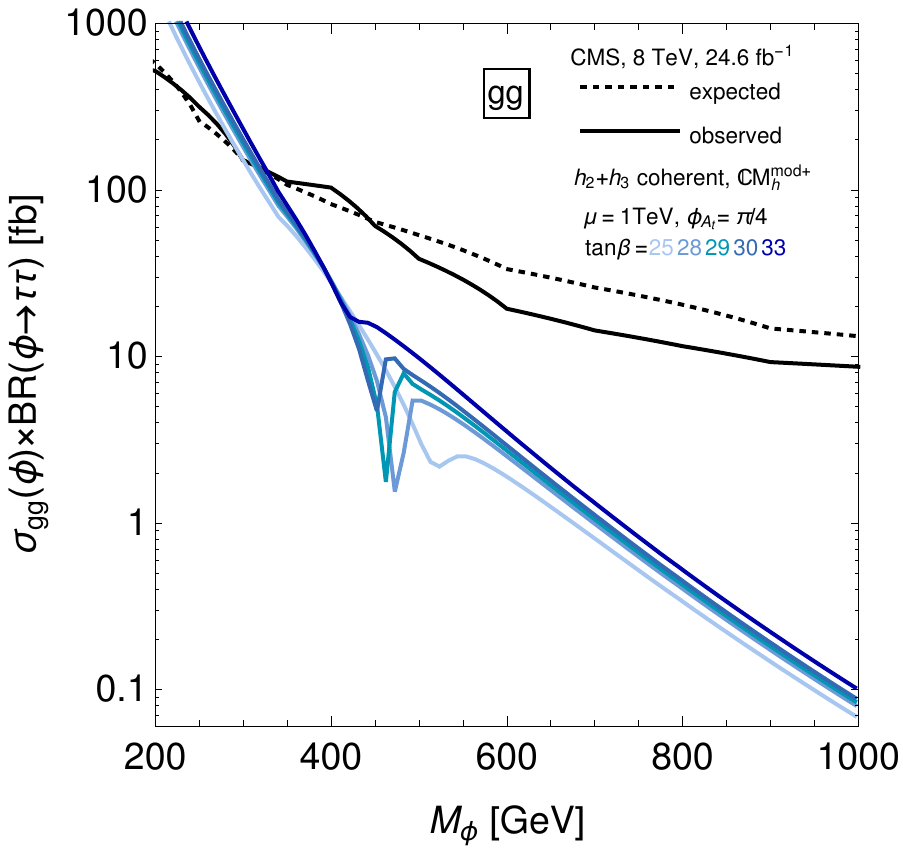}\label{fig:ggtb}}
  \caption{Comparison of the coherent $h_2$ and $h_3$ 
Higgs cross sections times branching ratio into $\tau^+\tau^-$ (including 
the interference contribution) 
in the $\mathds{C}\mhmod$ scenario with $\pat=\pi/4$ for
different values of
$\tb$ with $95\%$ CL exclusion bounds obtained by CMS at at $8\tev$.
See the respective colour coding for $\tb=25,28,29,30,33$. 
The displayed exclusion bounds are the same as in Fig.\,\ref{fig:InCohMphi}.
In (a) the production via $\bb$ is shown, in the (b) production via
$gg$.}
  \label{fig:CohTB}
 \end{center}
\end{figure}

In order to display the behaviour in the resonance region in more detail, 
Fig.\,\ref{fig:CohTB} shows the results for the coherent cross sections with
$\pat=\pi/4$ (including the interference contribution) 
for different values of $\tb$ across the resonance region. 
Both for the $\bb$ and $gg$ production modes, the most pronounced 
interference effect occurs at about $\tb=29$. The interference effect
is similarly large for $\Delta\tb = \pm 1$, while it can be seen to be
largely washed out and significantly shifted towards smaller (larger) values
of $\mphi$ for $\Delta\tb = + 4$ ($\Delta\tb = - 4$) in this example. As
before, the sensitivity of the $gg$ production mode is much smaller in this
parameter region, which gives rise to the fact that the main modification of
the cross section caused by the interference contribution occurs in the
region much below the excluded cross section. For the 
$\bb$ production mode, on the other hand, the interference contribution can
reduce the cross section to a level below the exclusion limit within the
resonance region.

In should be noted in this context that the comparison between the predicted
cross section and the observed exclusion limit is of course affected by
theoretical (see the discussion in Ref.\,\cite{deFlorian:2016spz}) 
and experimental\,\cite{CMS-PAS-HIG-13-021}
uncertainties. Since the main purpose of our
discussion has been the illustration of the qualitative features of the
$3\times 3$ mixing and the interference contribution, we do not address the
issue of uncertainties in detail here. 
Assuming 
comparable experimental uncertainties in the scenario considered 
in this work and in the standard $\cmhmod$ scenario analysed in 
Refs.\,\cite{CMS-PAS-HIG-13-021,Khachatryan:2014wca},
the impact of the mixing and interference effects 
extends beyond the experimental uncertainty band. 
Concerning the theoretical uncertainty of the
interference contribution, it should be noted that our approach outlined in 
Sect.\,\ref{sect:higherorder} involves a systematic treatment of the finite
widths of the resonantly produced Higgs bosons, which is crucial for a
proper treatment of the interference effects. For a more detailed
discussion of the uncertainties
related to the incorporation of the
interference contribution we refer to
Ref.\,\cite{IntCalc:InProgress}.
\subsection{Exclusion bounds with mixing and interference}
\label{sect:HB}

We now turn to an analysis of the exclusion limits in the $(\mhp,\tanb)$
parameter plane of the $\cmhmod$ scenario using 
the program 
\texttt{HiggsBounds-4.2.0}\,\cite{\HBlabel}, in which
comprehensive
information about limits from a variety of searches
at LEP, the Tevatron and Run~1 of the LHC
in different channels is
implemented.
Among the channels that are most important for constraining
the $\cmhmod$ scenario are
the ATLAS and CMS searches for neutral heavy Higgs bosons decaying 
to $\tau\tau$~\cite{Khachatryan:2014wca,Aad:2014vgg} and
to $ZZ(\rightarrow 4l)$~\cite{CMS:xwa,ATLAS:2013nma},
as well as analyses of the neutral Higgs boson at a mass of $125\gev$~\cite{CMS:aya,ATLAS:2012ae}
and searches for charged Higgs bosons~\cite{CMS:2014cdp,ATLAS:2014zha}.
For each parameter point,
\texttt{HiggsBounds} determines the search channel at LEP, 
Tevatron or the LHC with the highest expected sensitivity for an exclusion. 
For this channel the observed limit is confronted with the predicted cross
section in order to determine whether the considered parameter point is 
excluded at the 95\%\,CL.
For the channels
$\sigma_{\bb}(\phi)\,\br(\phi\rightarrow \tpm)$ and
$\sigma_{gg}(\phi)\,\br(\phi\rightarrow \tpm)$ 
the 2-dimensional likelihood
information provided by CMS\,\cite{CMS:2015mca,Khachatryan:2014wca} has 
been implemented%
\footnote{The recent \texttt{HiggsBounds-5-$\beta$} update~\cite{HB:5beta} which includes also Higgs search results from Run~2 of the LHC will be used in Ref.\,\cite{IntCalc:InProgress}.}.
For the incorporation of interference effects we have rescaled the ratio of
production cross sections used as input for 
\texttt{HiggsBounds} as described in (\ref{eq:relint1}) and
(\ref{eq:relint2}).

\begin{figure}[ht!]
 \begin{center}
   \subfigure[]{\includegraphics[width=0.46\textwidth]{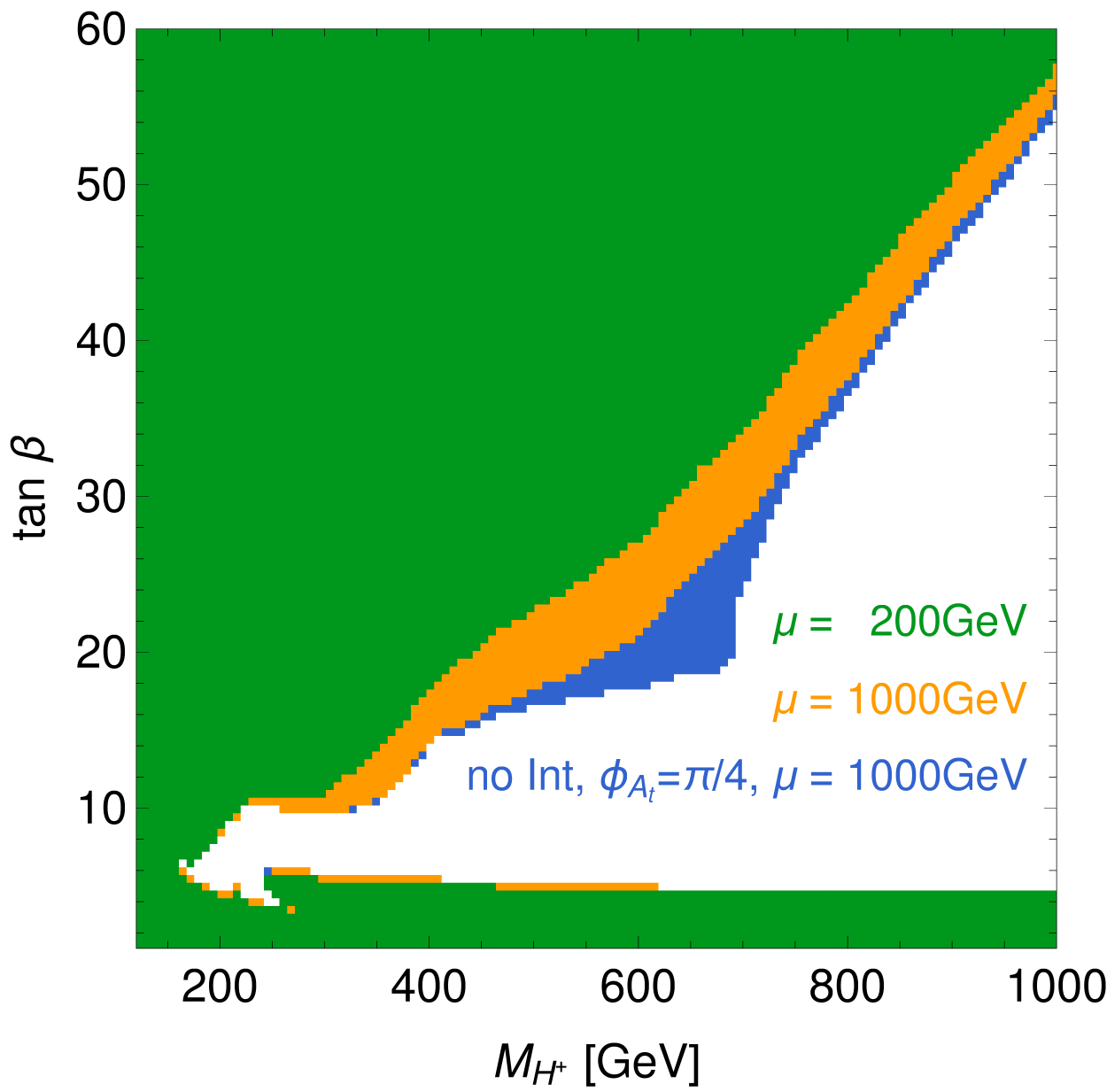} \label{fig:HBmix3}}
  \subfigure[]{\includegraphics[width=0.46\textwidth]{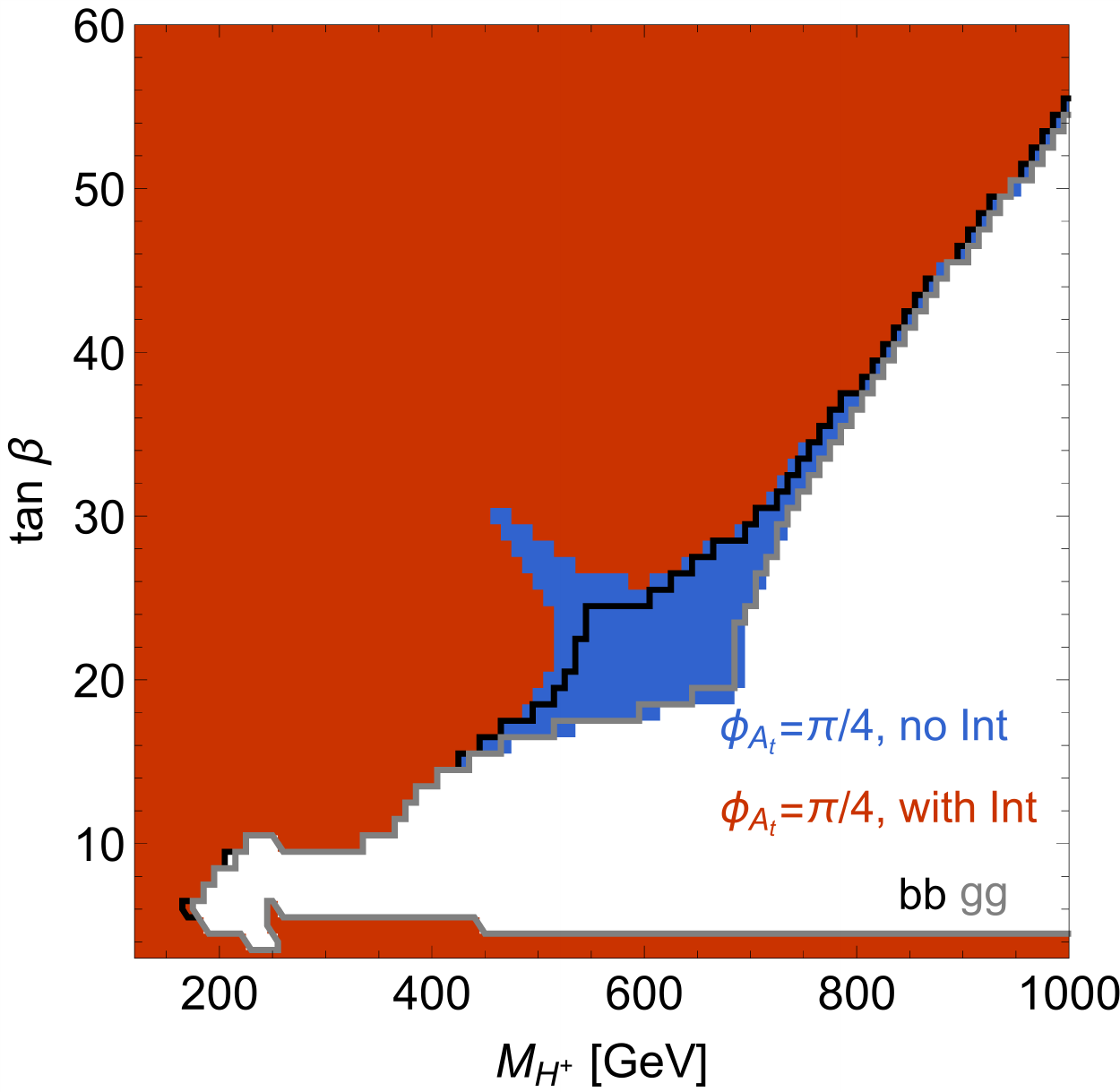}  \label{fig:HBbbgg}}
  \caption{Exclusion bounds in the $(\mhp,\tanb)$ plane of the MSSM 
obtained with \texttt{HiggsBounds}.
  (a) $\mhmod$ scenario with real parameters for $\mu=200\gev$ (green) and
$\mu=1\tev$ (orange); exclusion bound that would be obtained in the 
$\cmhmod$ scenario with $\pat=\pi/4$ for the 
incoherent sum of cross sections with $3\times 3$ mixing but without
the interference contribution (blue). 
  (b) $\cmhmod$ scenario with $3\times3$ mixing but without 
the interference contribution (blue, same as in Fig.\,\ref{fig:HBmix3}), 
including the
interference contribution in both the $\bb$ and the $gg$ processes (red),
including the interference contribution only in $\bb$ (black line) or
only in $gg$ (grey line).}
  \label{fig:HB}
 \end{center}
\end{figure}

In order to disentangle the interference effects from other effects due to
the modified scenario, the result for the $\cmhmod$ scenario with
$\pat=\pi/4$ and $\mu=1000\gev$ is shown in 
Fig.\,\ref{fig:HB} 
together with the familiar exclusion
region in the well-known $\mhmod$ scenario\,\cite{Carena:2013ytb} with real
parameters (in particular $\pat = 0$)
and the default value of $\mu=200\gev$. The latter exclusion region
is indicated in green in Fig.\,\ref{fig:HBmix3},
where the unexcluded parameter region in this type of plot is often
denoted as the ``LHC wedge'' region.
The corresponding exclusion region for 
$\mu=1000\gev$ is shown in orange. The larger value of $\mu$ yields 
stronger exclusion limits since in this case the
decay channel of a heavy Higgs boson into higgsino-like neutralinos and
charginos is kinematically closed, resulting in an enhancement of the 
branching ratio into $\tau^{+}\tau^{-}$.
The excluded region that one would obtain for the incoherent sum of
cross sections
of the $\cmhmod$ scenario with $\pat=\pi/4$ and $\mu=1000\gev$ 
($3\times3$ mixing, no interference contribution taken into account)
is displayed in blue in the same plot. We find that the 
incoherent cross section
of the $\cmhmod$ scenario with $\pat = \pi/4$ would result in a stronger
exclusion than the corresponding scenario with real parameters and the same
value of $\mu$
because of
the mixing-enhanced production cross sections
as discussed
in Sects.\,\ref{sect:mixenhance} and \ref{sect:InCoh}.

In Fig.\,\ref{fig:HBbbgg} the impact of the interference contributions on
the excluded regions in the $\cmhmod$ scenario is investigated.
The blue area (the same
as in Fig.\,\ref{fig:HBmix3}) corresponds to the incoherent sum of 
cross sections without 
the interference contribution. In contrast, taking the interference
contribution into
account in both the $\bb$- and the $gg$-initiated processes leads to the
weaker exclusion limit given by the red area, as a consequence of the
destructive interference effect as discussed in 
Sect.\,\ref{sect:InCoh}. The black line 
indicates the boundary of the
excluded region that one would obtain if the interference contribution were
taken into account only for the 
$\bb$-initiated process, while the grey line 
indicates the boundary of the
excluded region that one would obtain if the interference contribution were
taken into account only for the 
$gg$-initiated process.

The features visible in Fig.\,\ref{fig:HBbbgg} can be related to the results
of Sect.\,\ref{sect:InCoh}, since each horizontal line of constant $\tb$
in Fig.\,\ref{fig:HBbbgg} approximately 
corresponds to one plot as in Fig.\,\ref{fig:InCohMphi} for the same $\tb$,
separately for each process. It should be noted in this context, however,
that in 
Fig.\,\ref{fig:HBbbgg} also contributions from the lightest neutral Higgs
boson of the MSSM are taken into account, and the experimental information
implemented in \texttt{HiggsBounds} is employed, consisting in particular of
a 2-dimensional likelihood distribution\,\cite{Bechtle:2015pma} 
for the production via $\bb$ and
$gg$, instead of a single cross-section limit. Furthermore, the quantity on
the x-axis of Fig.\,\ref{fig:HBbbgg} is $\mhp$, while the results in 
Fig.\,\ref{fig:InCohMphi} and Fig.\,\ref{fig:CohTB} are expressed in terms
of the mass of a (single) neutral resonance, $\mphi$. While the interference
contributions have little impact on the exclusion region for $\tb$ values
below about 15, their effect is clearly visible for $\tb \gsim 15$ up to the
highest values of $\tb$ shown in the plot. The unexcluded ``fjord'' between
about ($\mhp = 480\gev, \tb = 29$) and 
($\mhp = 600\gev, \tb = 20$) is a consequence of the destructive
interference contributions in the resonance region discussed in 
Sect.\,\ref{sect:InCoh}. It is interesting to note that the full ``fjord''
only occurs as a consequence of incorporating the interference contributions
in both the $\bb$- and $gg$-initiated processes. If the 
interference contribution were incorporated only for the $\bb$-initiated process
but neglected for the $gg$-initiated process the resulting exclusion region
would have the ``bay'' shape indicated by the black line in 
Fig.\,\ref{fig:HBbbgg}. The ``fjord'' between $25 \lsim \tb \lsim 29$ 
would appear to be excluded in this case.
The reason for this relatively large impact of the $gg$-initiated process in
a parameter region where it is sub-dominant (see 
Fig.\,\ref{fig:InCohMphi} and Fig.\,\ref{fig:CohTB}) is mainly due to the
enhancement of the $gg$-initiated process in the resonance region that would
occur 
for a non-zero phase  
if the interference contribution were neglected, see
Fig.\,\ref{fig:b}. While this parameter region is allowed by the 
\textit{coherent} $\bb$ cross section (see Fig.\,\ref{fig:a}), it would
appear to be excluded by the \textit{incoherent} enhanced $gg$ cross section 
in this case (see Fig.\,\ref{fig:b}). 
On the other hand, incorporating the interference contribution only for the
$gg$-initiated processes but not for the $\bb$-initiated one would only have
a minor impact, as can be seen by comparing the grey line in 
Fig.\,\ref{fig:HBbbgg} with the boundary of the blue area. 

\section{Conclusions}
\label{sect:concl}

In the present paper we have investigated how the limits from the LHC
searches for additional heavy SUSY Higgs bosons get modified if instead of
the assumption of $\cp$ conservation in the Higgs sector the general case 
of complex parameters giving rise to $\cp$-violating effects is 
taken into account. 
We have considered neutral Higgs boson production in gluon fusion and in
association with $\bb$, and the decay into a pair of $\tau$-leptons in an 
extension of the well-known $\mhmod$ 
benchmark scenario of the MSSM where $\cp$-violating effects
are induced by a non-zero phase of the trilinear coupling $A_t$, 
$\pat=\pi/4$.

While in the LHC searches for additional Higgs bosons in supersymmetric models 
so far it has been assumed that the contributions from $\cp$-even and $\cp$-odd 
states to the signal cross section can be added as an
incoherent sum, such an assumption is not valid in the general case where
$\cp$-violating effects in the Higgs sector are taken into account. 
In the Higgs sector of the MSSM, complex parameters 
enter via loop corrections, causing $\cp$-violating effects beyond the tree 
level. As a consequence, the $\cp$ eigenstates $h,H,A$ mix into the 
mass eigenstates $h_1, h_2, h_3$. The Higgs sectors of the NMSSM and of a
non-supersymmetric 2HDM are in general $\cp$-violating already at lowest
order. The most obvious realisation of an extended Higgs sector that is
compatible with the present experimental constraints 
is a scenario where one neutral
scalar is SM-like and has a mass of about 125~GeV, while the other neutral
Higgs bosons are significantly heavier. In such a decoupling-type scenario
the heavy neutral Higgs bosons of the extended Higgs sector are typically
nearly mass-degenerate. If $\cp$-violating effects are present,
potentially large mixing and interference effects between the heavy
neutral Higgs bosons occur as a generic feature of such extended Higgs
sectors. These kind of effects should therefore be taken into account in the
interpretation of the limits from the searches for additional neutral Higgs
bosons.

Using an approximation of the $(3\times 3)$ Higgs propagator matrix in terms
of wave function normalisation factors evaluated at the complex poles and
Breit--Wigner propagator factors, we have determined the relative
interference contributions to the full process of production and decay. We
have demonstrated how these interference contributions can be combined with
the results for the on-shell production and decay of individual Higgs bosons
in order to arrive at a prediction for the full process incorporating
interference effects. We have compared this result based on a coherent sum
of the relevant contributions with the corresponding incoherent sum that
would be obtained if interference contributions were omitted.

Furthermore,
we have analytically investigated the relative interference contribution 
with respect to its dependence on the couplings, wave function normalisation
factors, masses and widths. We have pointed out that besides the well-know 
criterion in terms of the mass
difference and sum of the total Higgs widths, the
occurrence of a sizeable $\cp$-violating interference effect is also
directly related to the presence of imaginary contributions in the
propagator matrix. Since the latter can be well approximated by the wave
function normalisation factors, we could derive a simple criterion
indicating whether interference effects are expected to be relevant.

In the considered scenario we have found that both the mixing effects and
the interference contributions are large in the resonance region where the
above criteria are fulfilled. In fact, if interference effects were
omitted, the mixing effect between the nearly mass-degenerate heavy neutral
Higgs bosons $h_2$ and $h_3$ would give rise to a significant enhancement of
the total cross section in this region. However, the large destructive
interference contribution, which in the considered scenario reaches a
minimum of $-97\%$ at $\mhp\simeq 480\gev$ and $\tanb\simeq 29$
in both production processes, overcompensates this effect.
The net effect is therefore a
large suppression of the total cross section times branching
ratio in the resonance region as compared to the case of real parameters.

We have studied the impact of those modifications from mixing and
interference effects in the presence of complex parameters on the limits
obtained from the LHC searches in comparison with the limits for the
$\cp$-conserving case of real parameters. For illustration we have first
investigated single cross section limits, followed by a more comprehensive
analysis in the $(\mhp,\tanb)$
parameter plane of the $\cmhmod$ scenario using the tool \texttt{HiggsBounds}.
As expected by our analytical investigation, we have found that both the
mixing enhancement and the relative effect of the destructive interference 
are similar for the two production modes of gluon fusion and 
associated production with $\bb$. For moderate and
large values of $\tanb$, the $\bb$ cross section is larger than the gluon
fusion one,
but both processes are relevant for the analysed parameters. We have
demonstrated that the $\cp$-violating mixing and interference effects 
cause a 
substantial shift of the exclusion bounds in the parameter space 
along 
the ``LHC wedge'' region in comparison to the limits that would be
obtained in the corresponding 
scenario with real parameters. We have shown that in order to obtain
those results it is important to incorporate the interference contribution
both for the $\bb$ associated production and the gluon fusion process. 
Our analysis has indicated that 
a considerable parameter region remains unexcluded by LHC searches from
Run~1 that would appear to be ruled out if interference effects were neglected.

The illustrative study carried out in this paper 
motivates the analysis of current and future
LHC results in scenarios with complex
parameters, taking interference effects into account.  
While in our qualitative investigation we have restricted
ourselves to the analysis of the experimental results that were obtained at
Run~1 of the LHC and we have used an approximate treatment of the cross
sections for the production processes, our results lay the foundation for a
more detailed phenomenological analysis incorporating
the latest experimental results and the most accurate theoretical predictions.
We defer such a more detailed investigation taking into account the latest
experimental results from Run~2 of the LHC, employing the recently released
tool \texttt{SusHiMi} for the accurate prediction of the Higgs production processes 
in gluon fusion and $\bb$ associated production, and studying the variation
of the phases of several 
relevant MSSM parameters, in particular the phase of the gluino mass parameter,
to future work\,\cite{IntCalc:InProgress}.


\vspace{1cm}

\section*{Acknowledgements}
We would like to thank Oscar St\aa l for his help with \texttt{HiggsBounds} and valuable suggestions, and Christian Veelken, Sven Heinemeyer, Shruti Patel and Stefan Liebler for useful discussions. E.F. thanks the DESY theory group where a large part of this work was done. The work of E.F. was partially funded by the German National Academic Foundation and by DESY. The work of G.W.\ is supported in part by 
the Collaborative Research Centre SFB~676 of the DFG, ``Particles, Strings and 
the Early Universe'', and by the European Commission through the ``HiggsTools'' 
Initial Training Network PITN-GA-2012-316704.

\bibliographystyle{utphys}
\bibliography{InterferenceThExp,PhDthesis,bbH,gNWALoop,mthesis_literature}
\markboth{}{}
\end{document}